\documentclass[deluxetables]{emulateapj}
\usepackage{apjfonts}
\usepackage{epsf}
\begin{document}
\slugcomment{}
\shortauthors{J. M. Miller et al.}
\shorttitle{GRS 1915$+$105 in a Seyfert 2-like State}

\title{An Obscured, Seyfert-2-like State of
  the Stellar-mass Black Hole GRS 1915$+$105 Caused by Failed Disk Winds}

\author{J.~M.~Miller\altaffilmark{1},
A.~Zoghbi\altaffilmark{1},
J.~Raymond\altaffilmark{2},
M.~Balakrishnan\altaffilmark{1},
L.~Brenneman\altaffilmark{2}  
E.~Cackett\altaffilmark{3},
P.~Draghis\altaffilmark{1},
A.~C.~Fabian\altaffilmark{4},
E.~Gallo\altaffilmark{1},
J.~Kaastra\altaffilmark{5,6}
T.~Kallman\altaffilmark{7},
E.~Kammoun\altaffilmark{1},
S.~E.~Motta\altaffilmark{8},
D.~Proga\altaffilmark{9},
M.~T.~Reynolds\altaffilmark{1},
N.~Trueba\altaffilmark{1}
}

\altaffiltext{1}{Department of Astronomy, University of Michigan, 1085
  South University Avenue, Ann Arbor, MI 48109-1107, USA,
  jonmm@umich.edu}

\altaffiltext{2}{Center for Astrophysics | Harvard \& Smithsonian, 60
  Garden Street, Cambridge, MA 02138, USA}

\altaffiltext{3}{Department of Physics \& Astronomy, Wayne State
  University, 666 West Hancock Street, Detroit, MI, 48201, USA}

\altaffiltext{4}{Institute of Astronomy, University of Cambridge,
  Madingley Road, Cambridge CB3 OHA, UK}

\altaffiltext{5}{SRON Netherlands Institute for Space Research,
  Sorbonnelaan 2, 3584 CA Utrecht, NL}

\altaffiltext{6}{Department of Physics and Astronomy, Universiteit
  Utrecht, PO Box 80000, 3508 TA Utrecht, NL}

\altaffiltext{7}{NASA Goddard Space Flight Center, Code 662,
  Greedbelt, MD 20771, USA}

\altaffiltext{8}{Department of Physics, Astrophysics, University of
  Oxford, Denys Wilkinson Building, Keble Road, OX1 3RH Oxford, UK}

\altaffiltext{9}{Department of Physics, University of Nevada, Las
  Vegas, Las Vegas, NV 89154, USA}

\begin{abstract}
We report on {\it Chandra} gratings spectra of the stellar-mass black
hole GRS 1915$+$105 obtained during a novel, highly obscured state.
As the source entered this state, a dense, massive accretion disk wind
was detected through strong absorption lines.  Photionization modeling
indicates that it must originate close to the central engine, orders
of magnitude from the outer accretion disk.  Strong, nearly sinusoidal
flux variability in this phase yielded a key insight: the wind is
blue-shifted when its column density is relatively low, but {\em
  red-shifted} as it approaches the Compton-thick threshold.  At no
point does the wind appear to achieve the local escape velocity; in
this sense, it is a ``failed wind.''  Later observations suggest that
the disk ultimately fails to keep even the central engine clear of
gas, leading to heavily obscured and Compton-thick states
characterized by very strong Fe~K emission lines.  Indeed, these later
spectra are successfully described using models developed for obscured
AGN.  We discuss our results in terms the remarkable similarity of GRS
1915$+$105 deep in its ``obscured state'' to Seyfert-2 and
Compton-thick AGN, and explore how our understanding of accretion and
obscuration in massive black holes is impacted by our observations.
\end{abstract}

\section{Introduction}
Efforts to unify different classes of AGN are often discussed as
positing that orientation is the overarching factor in determining the
appearance of various sources.  In fact, this is merely an abstraction
that was largely ruled out in classical treatments (e.g., Antonucci
1993, Urry \& Padovani 1995).  A more accurate distillation -- since
backed by countless observations -- is that geometry is a key factor
{\em at a given fraction of the Eddington luminosity}.  However, it is
difficult to unambiguously determine the role of the Eddington
fraction in shaping AGN appearances and phenomena: indirect
measurements suggest that a typical AGN lifetime (the time when the
AGN luminosity is above a specified minimum Eddington fraction) is at
least $10^{7}$~years (e.g., Martini \& Weinberg 2001, Marconi et
al.\ 2004).  By force, this also complicates efforts to separate the
phenomena common to accretion in a deep gravitational potential,
from those that are set by environmental factors far from the black
hole.

Stellar-mass black hole outbursts and multi-wavelength ``states'' span
several orders of magnitude in luminosity in just weeks and months
(for a review, see, e.g., Remillard \& McClintock 2006).  The promise
of such studies is that AGN evolution driven by variations in the
Eddington-scaled mass accretion rate, $\dot{m}$, may be revealed on
accessible time scales.  In this regard, GRS 1915$+$105 may be the key
stellar-mass black hole of the modern era.  Whereas the states
observed in this and other stellar-mass black holes may correspond to
different AGN classes in broad terms (e.g., Svoboda et al.\ 2017),
{\it specific} multi-wavelength behaviors and X-ray phenomena observed
in GRS 1915$+$105 have clear, mass--scaled counterparts in AGN.

Most importantly, cycles of X-ray flux dips and subsequent radio
flares seen in GRS 1915$+$105 are also observed in quasars, and the
relative period scales simply with mass (e.g., Mirabel \& Rodriguez
1998, Marscher et al.\ 2002).  The fastest disk winds and extreme disk
reflection observed in numerous AGN also have clear analogies in GRS
1915$+$105 (e.g., Tombesi et al.\ 2010, Miller et al.\ 2016, Zoghbi et
al.\ 2016).  The ``heartbeat'' oscillations occasionally observed in
GRS 1915$+$105 -- likely limit-cycle variations (e.g., Neilsen et
al.\ 2011, Zoghbi et al.\ 2016, Motta et al.\ 2020) -- and so-called
quasi-periodic eruptions (or, QPEs; see Miniutti et al.\ 2019,
Giustini et al.\ 2020) seen in some AGN may also arise from the same
underlying physical process.

Perhaps because black holes have particle-like qualities and are true
points within General Relativity, it is tempting to regard key
quantities like $\dot{m}$ and obscuration (as a proxy for orientation)
as fully orthogonal eigenvectors.  Nominally, $\dot{m}$ determines the
source luminosity, while the orientation of the source determines how
obscured it is.  This ethos may also be reinforced by the fact that
obscuration in some AGN is clearly due -- at least in part -- to dust
lanes in the host galaxy (e.g., NGC 4388, Pogge \& Martini 2002).  It
is also possible that the nature of a distant, pc-scale, obscuring
``torus'' are partly determined by larger environmental factors,
including the character of gas flows into the nucleus (see, e.g.,
Ramos Almeide \& Ricci 2017).

However, $\dot{m}$ and obscuration may be related in some
circumstances, potentially rendering obscuration a poor proxy for
orientation.  For instance, Compton-thick AGN are often associated
with recent mergers and rapid black hole accretion (e.g., Komossa et
al.\ 2003, Koss et al.\ 2018).  Recent mergers are relatively easy to
identify, but the situation would be more complicated if heavy
obscuration or even Compton-thick phases could manifest at much lower
$\dot{m}$.  So far, clear analogies between stellar-mass black holes
and AGN have mostly been limited to unobscured Type-I AGN.  Could
stellar-mass black holes also reveal connections between $\dot{m}$ and
obscuration that would impact our view of obscured Type-II AGN, and
AGN evolution?

In mid-2018, X-ray monitoring began to suggest that GRS 1915$+$105 was
entering an unusual low-flux state (Negoro et al.\ 2018).
By May 2019, the observed flux was just 50 mCrab, nearly an
order of magnitude below flux levels commonly observed in the
``low/hard'' state (Miller et al.\ 2019a).  Continued {\it
  Swift} monitoring revealed that the accretion flow in GRS 1915$+$105
was occasionally Compton-thick (Miller et al.\ 2019,
Balakrishnan et al.\ 2019) and that obscuration was a key
factor in the diminished X-ray flux.  Strong flaring in radio bands
also indicated that $\dot{m}$ may still be relatively high (Motta et
al.\ 2019).

Monitoring has revealed that the ``obscured state'' of GRS 1915$+$105
is not the mass-scaled equivalent of a brief obscuration event similar
to a ``changing-look'' event in an AGN (see, e.g., Matt et al.\ 2003;
however, also see Runnoe et al.\ 2016).  Rather, internal obscuration
that is generally well above $N_{H} \geq 10^{23}~{\rm cm}^{-2}$ and
sometimes Compton-thick has endured for over a year (Balakrishnan et
al.\ 2020, in prep.).  This duration likely represents at least one viscous time
scale through the entire disk in GRS~1915$+$105.  In this sense, it is
not a momentary change in the accretion flow geometry, nor a
phenomenon exclusively linked to the corona or jet production.
Rather, it is more likely to be an entirely new accetion state.

Owing to its ongoing importance across the black hole mass scale, the
binary parameters and components of GRS 1915$+$105 have been studied
extensively.  The black hole primary and its K-type giant companion
have an orbital period of ${\rm P} = 33.85\pm 0.16$~days (Steeghs et
al.\ 2013).  Radio imaging of approaching and receding knots in the
relativistic jet have suggested that the inner disk is viewed at an
inclination of $\theta = 66\pm 2$~degrees (Fender et al.\ 1999).  The
best measurement of the black hole mass is $M_{BH}
=12.4^{+2.0}_{-1.8}~M_{\odot}$, for a parallax distance of $d =
8.6^{+2.0}_{-1.6}$~kpc (Reid et al.\ 2014).  The spin of the black
hole itself is nearly maximal, with $a = 0.98$ (where $a = cJ/GM^{2}$;
McClintock et al.\ 2006, Miller et al.\ 2013).

In this paper, we report on an analysis of three {\it Chandra}
gratings observations of GRS 1915$+$105.  The first of these was
obtained as the source entered the obscured state; the latter two were
obtained in the midst of this state.  Section 2 describes the
observations and how they were reduced.  Section 3 describes our
analysis method and results.  In Section 4, we discuss the
implications of our results for GRS 1915$+$105, the nature and origin
of this new spectral state, and our understanding of Compton-thick and
heavily obscured AGN.

\section{Observations and Data Reduction}
Table 1 lists the start times and net exposure obtained (after
filtering) for observations ({\it Chandra} ObsIDs) 22213, 22885, and
22886.  All observations used the ``faint'' ACIS imaging mode.  ObsID
22213 utilized a 350-row sub-array to limit photon pile-up; ObsIDs
22885 and 22886 utilized 512-row sub-arrays.

Each observation was reduced using CIAO version 4.11, and the
corresponding calibration files.  We extracted the first- and
third-order spectra from each observation.  The standard ``mkgrmf''
and ``fullgarf'' routines were used to construct corresponding
response files.  Spectra and responses from the opposing orders were
added using the ``combine\_grating\_spectra'' tool.  None of the
third-order spectra obtained enough photons to conduct a sensible
analysis.  The MEG spectra would nominally be more sensitive than the
HEG at low energy, but owing to the combination of line-of-sight and
internal obscuration, the MEG spectra are uniformly less sensitive
than the HEG spectra.  Therefore, our analysis focused on the combined
first-order HEG spectrum from each observation.

Particularly in ObsID 22213, the zeroth order image of the source is
adversely affected by photon pile-up, so we extracted light curves
from the first-order HEG and MEG events using the tool ``dmextract''.
This observation shows strong, nearly sinusoidal variations; these are
discussed at length in the next section.  The light curve of ObsID
22885 shows low-level variability, typical of that observed in
accreting sources.  ObsID 22886 shows a shallow rising trend in the
count rate over the course of the observation.

\section{Analysis \& Results}
We analyzed time-averaged and time-selected spectra using SPEX
(Kaastra et al.\ 1996) and XSPEC (Arnaud 1996).  In all cases, we used
Cash statistics (Cash 1979) to assess the goodness of each fit and the
significance of specific features.  Initially, models were fit with
standard weighting based on the number of counts per bin; once a good
fit was found, refinements were made using ``model'' weighting.  The
latter step avoids biases caused by artificially large errors on the
zero-flux side of an emission or absorption spectrum.  All errors
quoted in this work reflect the value of a given parameter at the
boundary of its $1\sigma$ confidence interval.

A preliminary examination of the spectra is shown in Figure 1.  For
comparison, we show a spectrum of GRS 1915$+$105 in a disk--dominated
``high/soft'' state, exhibiting very strong and complex wind with
accretion disk P Cygni profiles (Miller et al.\ 2015, 2016).  ObsID
22213 was clearly obtained at a flux that is an order of magnitude
lower; however, the spectrum is still dominated by a number of strong
absorption lines.  Finally, ObsID 22885 is plotted.  The continuum
level at 5~keV is over two orders of magnitude below the soft state,
and the continuum level at 2~keV is at least three orders of magnitude
lower.  The spectrum is dominated by narrow Fe K emission lines that
are several times stronger than the local continuum, signaling that
most of the flux required to excite the lines is obscured.  This is
broadly consistent with the Compton-thick obscuration that is observed
in some AGN (see, e.g., Kallman et al.\ 2014, La Massa et al.\ 2017,
Kammoun et al.\ 2019, 2020).

\subsection{Entering obscuration: ObsID 22213}
Figure 2 shows the time-averaged spectrum of ObsID 22213, as
represented within SPEX.  The spectrum is marked by He-like and H-like
absorption lines from Si, S, Ar, Ca, and Fe.  The clearest emission
line in the spectrum is seen close to 6.40~keV, consistent with an
Fe~K$\alpha$ fluorescence line from dense neutral material or emission
from more diffuse gas with a low ionization.  The sensitivity of the
spectrum is limited below 3~keV, simply owing to Galactic absorption
along the line of sight.  For this reason, we chose to focus on the
3.0--10.0~keV band in our spectral fits to ObsID 22213.

The strongest absorption feature in the entire spectrum is a broad
trough between roughly 6.4 keV (Fe I-XVII) and 6.7~keV (He-like Fe
XXV).  This feature indicates that a broad range of charge states
contribute to the absorption spectrum.  Blending due to a range of
velocity shifts may contribute subtlely.  Re-emission from such a
trough is expected to be weak owing to the low fluorescence yield of
the charge states involved, and this is consistent with the observed
spectrum.  Initial modeling confirmed that it is not possible to fit
this absorption trough {\rm and} the H-like Fe~XXVI line close to
6.97~keV using a single ionization zone (characterized by a single
ionization parameter, column density, and velocity).  Instead, at
least two distinct absorption zones are required by the data.

Within SPEX, the ``pion'' model (e.g., Mehdipour et al.\ 2015) affords
the chance to layer absorption zones so that an exterior zone sees the
continuum {\em after} it has been modified by an interior zone (see,
e.g., Trueba et al.\ 2019).  We therefore proceeded to model the
time-averaged spectrum from ObsID 22213 using SPEX.  We employed
``optimal binning'' in order to maximize the signal in the spectrum
(Kaastra \& Bleeker 2016).

The model we fit consisted of neutral line-of-sight absorption in the
Milky Way, acting on two ``pion'' zones covering a power-law continuum
and reflection.  In SPEX parlance, the model can be written:
$absm\times pion\times pion\times refl$.  The key model details and
parameters are as follows:\\

\noindent{\em Line of sight absorption}: The $absm$ model is the
Morrison \& McCammon (1983) absorption model (comparable to ``phabs'' in
XSPEC).  It is characterized by the equivalent neutral hydrogen column
density, $N_{H}$, and the covering factor, $f$.  We fixed the column
density at a value of $N_{H} = 5\times 10^{22}~{\rm cm}^{-2}$ (e.g.,
Miller et al.\ 2016, Zoghbi et al.\ 2016), and the covering factor to
unity (full covering).\\

\noindent{\em Photoionized absorption}: Each ``pion'' component is a
self-consistent photoionization model that adjusts to changes in the
incident spectrum as the minimization proceeds.  In both ``pion''
layers, the equivalent neutral hydrogen column density ($N_{H}$), gas
ionization (log$\xi$), rms velocity broadening ($\sigma$), and overall
velocity shift were allowed to vary.  The covering factor did not vary
significantly from unity in exploratory fits, so this value was then
fixed.  There is only weak evidence of re-emission from the absorbing
gas in the spectrum, so the separate covering factor for emission was
set to zero.  This is merely a simplifying assumption as the absorbing
gas must also emit and can potentially even reveal the gas location
(see, e.g., Miller et al.\ 2015, 2016; Trueba et al.\ 2019).  All of
the elemental abundances within ``pion'' were set to the solar
values.\\

\noindent{\em Reflection}: The ``refl'' model in SPEX is based on the
calculations of Magdziarz \& Zdziarski (1995).  The ``refl'' model has
15 total parameters, but a number of these are flags rather than true
variables.  We allowed the flux normalization, power-law index
($\Gamma$), the cosine of the inclination angle, and the reflection
``scale'' to vary (the ``scale'' is the relative strength of direct
and reflected emission, similar to the ``reflection fraction'' in
other implementations).  We froze the iron abundance at unity, the
power-law cut-off energy at $E_{cut} = 30$~keV (see, e.g., Blum et
al.\ 2009, Miller et al.\ 2013; however, also see Zdziarski et
al.\ 2005), and the reflector ionization at $\xi = 0$.  The ``refl''
model includes an internal blurring function to translate from the
fluid frame to the observed frame, characterized in terms of an inner
blurring radius ($r_{1}$), and outer blurring radius ($r_{2}$), and an
emissivity index ($q$, where $J \propto r^{-q}$).  We fixed the
emissivity to the Euclidian geometric value of $q=3$ (appropriate far
from the black hole; e.g., Wilkins \& Fabian 2012), and fixed $r_{2}$
to an arbitrarily large value ($r_{2} = 10^{6}~GM/c^{2}$).  The inner
disk radius, $r_{1}$, was allowed to vary.\\

Figure 3 shows the results of fits to the time-averaged spectrum with
this model, and Table 2 lists the resulting fit parameters and
$1\sigma$ errors.  This model provides a good description of the
time-averaged spectrum: $C = 1148.8$ for $\nu = 1090$ degrees of
freedom.  The most important outcomes from this model for the
time-averaged spectrum are: (1) the strong wind is likely composed of
two relatively slow components with distinct properties, and (2)
the wind likely originates within a small distance from the black
hole.  

The inner ``pion'' zone is highly ionized, with log$\xi =
4.2^{+0.1}_{-0.3}$.  It has a projected blue-shift of $v = -350\pm
70~{\rm km}~{\rm s}^{-1}$, and a column density of $N_{H} =
6.5^{+0.1}_{-0.2} \times 10^{23}~{\rm cm}^{-2}$.  An upper limit on
the wind launching radius can be obtained by recalling that $N_{H} =
n\Delta r$ and assuming that $\Delta r \simeq r$ so that $r \leq
L/N_{H}\xi$.  The best-fit model for the spectrum measures $L =
1\times 10^{38}~{\rm erg}~{\rm s}^{-1}$ (0.5-30~keV), which then gives
$r \leq 9.7\times 10^{9}~{\rm cm}$, or $r \leq 5.2\times
10^{3}~GM/c^{2}$.

The outer ``pion'' zone is quite distinct from the inner zone.  At
log$\xi = 3.0\pm 0.1$, it is an order of magnitude less ionized than
the inner zone.  It has a more modest column density of $N_{H} =
2.8^{+0.5}_{-0.4}\times 10^{23}~{\rm cm}^{-2}$, and its outflow
velocity of $v = -50^{+40}_{-20}~{\rm km}~{\rm s}^{-1}$ is comparable
to the uncertainty in the HETG wavelength scale (e.g., Ishibashi et
al.\ 2006).  Therefore, it is not clear that the outer zone is
outflowing.  In estimating an upper limit for the radius at which gas
in the outer pion zone is observed, it is important to use the {\em
  effective} luminosity ($L_{eff} \simeq 4.6\times 10^{37}~{\rm
  erg}~{\rm s}^{-1}$); this is the luminosity of the central engine
after filtering through the inner zone.  This gives $r \leq 1.6\times
10^{11}~{\rm cm}$, or $r \leq 8.8\times 10^{4}~GM/c^{2}$.

The reflection spectrum indicates weak blurring, driven by mild
broadening of the Fe~K$\alpha$ emission line (see Figure 3).
The best-fit blurring radius is $r = 2.2^{+7.8}_{-1.3} \times
10^{3}~GM/c^{2}$.  This radius is unlikely to represent the innermost
extent of the accretion disk, since prior studies have found that the
disk in GRS 1915$+$105 can extend to the ISCO at similar Eddington
fractions (e.g., Miller et al.\ 2013).  It is more likely that the
reflector represents dense, low-ionization, and potentially even
Compton-thick portions of a complex and stratified wind.

Since the observed Fe~K emission line is neutral (or, low-ionization),
the inner zone of the wind is likely contained within this reflection
radius.  Indeed, the radius constraint from the reflection spectrum
and the upper limit on the wind launching radius are in broad
agreement.  The escape velocity from $r = 2.2\times 10^{3}~GM/c^{2}$
is $v_{esc} = 9.0\times 10^{3}~{\rm km}~{\rm s}^{-1}$.  Even if the
gas is launched nearly vertically from the disk while retaining all of
its local Keplerian velocity ($v_{Kep} = 6.4\times 10^{3}~{\rm
  km}~{\rm s}^{-1}$), the gas is nominally unable to escape the
system.  The outer zone is not clearly outflowing.  It is immediately
apparent, then, that these dense failed wind components may eventually
build-up the obscuration that defines this new state of GRS
1915$+$105.

If we regard the reflection radius as an independent upper limit on at
least the inner zone of the wind, it can be used to derive an estimate
of the density and filling factor of this zone.  Transforming the
gravitational units of ``refl'' into physical units and writing $n = L
/ \xi r^{2}$, the inner ``pion'' zone has a density of $n \simeq 3.8
\times 10^{14}~{\rm cm}^{-3}$.  Taking $f = \Delta r/r = N_{H} / n r$,
this implies that $f \geq 0.4$.  It is notable that the inner zone of
this wind is broadly similar to that observed in steady soft states of
GRS 1915$+$105, in terms of its density and launching radius.
However, it is much slower than the zones observed in soft states
(e.g., Neilsen \& Lee 2009, Ueda et al.\ 2009, Miller et al.\ 2015,
2016), suggesting a much smaller driving force.

The mass ``outflow'' rate of each zone is given by $\dot{M}_{wind} =
\Omega \mu m_{p} f n r^{2} v$ or $\dot{M}_{wind} = \Omega \mu m_{p} f
(L/\xi) v$, where $\Omega/4\pi$ is the covering factor, $\mu$ is the
mean atomic weight (assumed to be $\mu = 1.23$ as per material with
solar abundances), $m_{p}$ is the mass of the proton, $f$ is the
volume filling factor, $L$ is the radiative luminosity, $\xi$ is the
ionization parameter, and $v$ is the outflow velocity.  The kinetic
power of each zone is then just $L_{kin} = 0.5 \dot{M}_{wind} v^{2}$.
Assuming $\Omega = 2\pi$ and $f = 0.4$ will allow for a reasonable
estimate, and one from which it is easy to scale if better information
becomes available (note that other recent studies suggest high filling
factors; see Trueba et al.\ 2019, Balakrishnan et al.\ 2020, in prep.).  For the
inner ``pion'' zone, we estimate that $\dot{M}_{wind} \simeq 1.1
\times 10^{18}~{\rm g}~{\rm s}^{-1} = 1.7 \times
10^{-8}~M_{\odot}~{\rm yr}^{-1}$.  This is roughly equal to the
implied mass accretion rate (for an efficiency of $\eta = 0.1$):
$\dot{M}_{acc} \simeq 1.1\times 10^{18}~{\rm g}~{\rm s}^{-1} =
1.7\times 10^{-8}~ M_{\odot}~{\rm yr}^{-1}$.  This implies that half
of the gas available for rapid accretion is at least delayed in a
flow, even if it is not expelled to infinity.  The kinetic power in
the inner zone is $L_{kin} \simeq 6.7\times 10^{32}~{\rm erg}~{\rm
  s}^{-1}$.  This power is a small fraction of the radiative
luminosity, and the result of a low velocity; both factors indicate
that the radiation and gas are not coupled in a manner consistent with
super-Eddington accretion.

\subsection{Variations within ObsID 22213}
Figure 4 shows the light curve of ObsID 22213, with time bins of
$\Delta t = 50$~s.  Strong variations are clearly evident,
sometimes reaching an amplitude of $\pm$50\%.  These waves are not
exactly sinusoidal, and the variations appear to change slightly in both
frequency and phase.  Defining even a quasi-period with a relatively
small number of cycles is problematic, but the variations have a
typical ``period'' of ${\rm P} \simeq 1250$~s.

Several arguments clearly indicate that the observed variations are
astrophysical, not instrumental.  The same variations are seen when
different bin sizes are adopted, so they are not the spurious result
of an unfortunate binning.  {\it Chandra} executes a dither pattern
with a peak-to-peak span of 16 arc seconds, with nominal periods of
707~s and 1000~s in the X and Y coordinates of the ACIS pixels,
respectively.  An astrophysical origin for the variations is also
consistent with the fact that the dispersed light curves of other
comparably bright sources (Seyferts, and some X-ray binaries) do {\em
  not} show similar variations.  Finally, light curves of
GRS~1915$+$105 from the {\it Neil Gehrels Swift Observatory} taken
contemporaneously {\it do} show comparable variations (Miller et
al.\ 2019).

We modeled the dispersed light curve as a Gaussian Process (GP;
Rasmussen \& Williams 2005) in order to define oscillation phases relative
to troughs and crests, and to extract phase-selected spectra.
Specifically, we used the formalism where the covariance function is
represented by a mixture of exponentials (Foreman-Mackey et
al.\ 2017).  We use two covariance kernels: one to model the
periodicity, and one to model any remaining red noise in the light
curve.  The GP models the periodicity without any prior assumption
about the waveform shape.  The resulting light curve model is also
shown in Figure 4.  Once a model is obtained by maximizing its
likelihood, the locations of the peaks and troughs are used to
phase-tag the time bins, allowing them to be grouped by phase,
producing good time intervals (GTIs) of the desired oscillation
phases.

These GTIs were subsequently used to extract spectra in the same
manner as the time-averaged spectrum detailed in the prior subsection.
An oscillation phase of $\phi=0$ represents the crest of the nearly
sinusoidal wave; we extracted spectra from eight phase bins.  Note
that the running mean of the light curve varies, and the amplitude of
the variability also changes.  This means that, depending on the
spectrum of the central engine and the obscuration, the relative
fluxes in the oscillation phase bins cannot be expected to follow a
simple pattern.  Rather, our procedure treats the phase with respect
to crests and troughs as meaningful, and examines the spectral
variations that result.

The results of fits to the oscillation phase-selected spectra are
listed in Table 2 and shown in Figure 5.  As per the time-averaged
spectrum, the phase-selected spectra were fit over the 3--10~keV band
in SPEX using Cash statistics, and binned using the ``optimal
binning'' scheme of Kaastra \& Bleeker (2016).  The same model that
was applied to the time-averaged spectrum was applied to the
phase-selected spectra, with a small number of simplifying
assumptions: (1) the velocity broadening of each absorption zone was
fixed to the best-fit value measured in the time-averaged spectrum,
(2) the velocity of the outer zone was fixed to zero, and (3) the
inner radius and inclination of the reflector were fixed to the
best-fit values measured in the time-averaged spectrum.  These steps
were necessary because of the reduced sensitivity in the
phase-selected spectra (note that each phase represents 3.8~ks of
exposure).

Figure 6 plots the column density and ionization of the inner and
outer absorption zone, the velocity shift of the inner zone, and the
properties of the reflector as a function of phase.  These are the
independent variables that can be measured directly from the spectrum;
other quantities of interest (e.g., $\dot{M}$, $L_{kin}$, $r$) depend
on these parameters.  The variations in the velocity of the inner
absorption zone are particularly striking, in that the gas is measured
to have a significant red-shift in the phases farthest from the crests
($\phi=0$).  This broadly coincides with the phases with the highest
column densities.

In Figure 7, the ionization and column density of each zone are
plotted versus the luminosity.  (In the case of the outer zone, the
parameters are plotted versus the effective luminosity, after losses
in the inner zone.)  Spearman's rank correlation coefficients and
false correlation probabilities were calculated for each pairing, and
the resulting values are noted in Figure 7.  The correlation
coefficients are high, but the compelling cases are only significant
at the 90--95\% level, partly owing to the small number of points.
Overall, the positive correlations indicate that more gas enters the
line of sight with increasing luminosity, and then becomes more
ionized.  This is suggestive of a wind that lifts the gas above the
disk, where it is then exposed to ionizing radiation.

Figure 8 plots the velocity of the inner absorption zone versus the
log of its ionization parameter, and versus its column density.  The
velocity is positively correlated -- but likely not significantly
correlated -- with the ionization of the gas.  In contrast, the
velocity of the gas is strongly correlated with the column density of
the gas ($\rho = 0.92$, P$=0.001$).  Most importantly, the flow has the
strongest blue-shift when the column is low, and becomes red-shifted
when the column is nearly Compton-thick.

The $v$ -- $N_{H}$ relationship for the inner zone may explain the
strong variations seen in ObsID 22213, and may also explain the origin
of the obscured state.  The wind may not ultimately escape to
infinity, but when its column is low the central engine is better able
to clear the gas from the vicinity of the black hole.  However, when
the column starts to become very high -- nearly Compton-thick -- the
central engine is no longer able to clear the gas.  Especially since
the gas is not clearly able to escape from the inferred
photoionization radius -- not even at its highest blue-shifts -- these
cycles are ultimately doomed to fail.  As the column density
accumulates, the data suggest that the flow will eventually envelop
the central engine.

It is appealing to ascribe strong flux variations to phenomena like
warps or precession; however, the data and the observed correlations
do not offer much support for ths explanation.  The observed
quasi-period of the variations in ObsID 22213, P$\simeq 1250$~s, is
the Keplerian orbital period at $r = 2.2\times 10^{4}~GM/c^{2}$.  The
inner absorption zone and reflector are likely interior to this
radius, and the outer zone is just external to this radius (though,
for a low filling factor, it may also be within it).  However, the
positive correlation between the obscuring column density and the
luminosity is difficult to explain in terms of a warp or precession.
Similarly, if the flux variations are caused by material passing
across our line of sight, the obscuring column density should be
evenly distributed with velocity (or, potentially clustered at
velocity extrema). 

\subsection{Deep obscuration: ObsIDs 22885 and 22886}
Figures 9--12 show the spectra obtained in ObsIDs 22885 and 22886, fit
with complementary models in XSPEC and SPEX.  In order to achieve the
best possible constraints on the subtantial internal column density
indicated in these spectra, the fitting range was extended down to
1.3~keV (the lower bound of the HEG).  The spectra are dominated by
narrow Fe K emission lines that are much stronger than the local
continuum component that excites them, indicating that most of the
continuum must be obscured.

On closer examination, the comparable strength of the He-like Fe XXV
forbidden, intercombination, and resonance lines indicates that the
ionized lines are produced via photoionization (e.g., Porquet \& Dubau
2000).  Simple calculations with CHIANTI (version 7.1; Dere et
al.\ 1997, Landi et al.\ 2013) confirm that photoionization must be
important.  A pure collisional ionization model with a temperature of
${\rm T} = 10^{7}$~K can approximately match the Fe XXV complex, but
it does not produce an Fe XXVI line.  A photoionized gas with a
similar temperature easily matches the Fe XXV and Fe XXVI lines.

We adopted an adaptive binning scheme to fit the data within SPEX.
The spectra were binned by a factor of 40 between 1.3--4.0 keV, by a
factor of 10 in the 4.0--6.0~keV band, by a factor of 2 in the
6.0--7.5~keV band, and finally by a factor of 20 in the 7.5--10.0~keV
band.  Prior to fitting in XSPEC, the spectra were grouped to require
at least 10 photons per bin.  These binning schemes differ, but each
has the effect of balancing resolution in the Fe~K band while also
extracting the most information possible from the broad continuum (in
order to accurately determine the internal absorbing column density).

Within XSPEC, the line of sight column density was modeled with
``phabs'', with $N_{H} = 5\times 10^{22}~{\rm cm}^{-2}$ fixed as per
fits to ObsID 22213.  We also modeled the internal column density with
``phabs'', acting on a cut-off power-law (with $E_{cut} = 30$~keV
fixed).  The power-law index was restricted to values common in the
low/hard state, $1.8\leq \Gamma \leq 2.2$.  We then accounted for the
strong neutral or low-ionization Fe K line at 6.4 keV using the
``xillver'' reflection model (Garcia et al.\ 2014), fixing log$\xi=0$.
The reflection model was included in a ``reflection only'' manner, to
prevent double-counting the continuum.  The power-law index and
cut-off energy in ``xillver'' were linked to the same parameters in
the cut-off power-law model.  The inclination was fixed at $\theta =
66^{\circ}$ (the fits were insensitive to this parameter.)  We fit the
Fe XXV and Fe XXVI lines using the ``photemis'' photoionized emission
model.  Like ``xillver,'' ``photemis'' is built from executions of
XSTAR (Kallman \& Bautista 2001), enabling self-consistency.
``Photemis'' assumes a $\Gamma = 2$ power-law input spectrum, which is
consistent with the range allowed for the cut-off power-law component.
We froze all abundances at their solar values, and we allowed the
``photemis'' ionization and flux normalization to float.  A series of
exploratory fits determined that a turbulent velocity parameter value
of $\sigma = 50~{\rm km}~{\rm s}^{-1}$ allowed the Fe XXV forbidden
line to be separated from the intercombination and resonance lines;
this value was fixed in all fits.  In XSPEC parlance, the model we
finally adopted can be written as follows: ``phabs*(phabs*cutoffpl +
xillver + photemis)''.

% now do spex

An analogous model was used within SPEX.  A line of sight column
density (via ``absm,'' with $N_{H} = 5\times 10^{22}~{\rm cm}^{-2}$
and $f_{cov} = 1$) asborbed all model components.  A second neutral
absorber acted on the direct continuum, but is allowed to only
partially cover the source.  The continuum spectrum was modeled in
terms of thermal Comptonization with ``comt'' rather than a power-law
because the photoionized emission component is particularly sensitive
to the divergence of a power-law at low energy.  We fixed the Wien
temperature and electron temperature at 0.2~keV and 120~keV,
respectively, and fit for the flux normalization and optical depth.
The Fe K line close to 6.4~keV was modeled using ``refl;'' as noted
previously, this is similar to ``pexrav'' in XSPEC.  We fit for the
component flux normalization (with a fixed power-law index of $\Gamma
= 1.8$), and the reflection scale factor.  Finally, via ``pion,'' we
fit for the column density, ionization, and emission scale factor
(essentially its normalization), and we froze the turbulent broadening
parameter at $\sigma = 50~{\rm km}~{\rm s}^{-1}$.  The total SPEX
model could be written as follows: $absm\times(absm\times comt + refl
+ pion)$.

The results of fits to the spectra obtained in ObsIDs 22285 and 22286
are presented in Table 3 and Figures 9--12.  The fits are good, but
not fomally acceptable.  In various trials, it is apparent that
additional photoionized emission components yield small improvements
in the fit, generally at the $3\sigma$ level of confidence.  Allowing
the abundance of Fe to be twice that of Ca and Si also yields minor
improvements.

The fits strongly indicate that ObsID 22285 suffered a much higher
internal column density than ObsID 22886.  Within XSPEC, a column of
$N_{H} = 2.2^{+0.9}_{-0.3}\times 10^{24}~{\rm cm}^{-2}$ is measured;
this model achieves a better goodness-of-fit statistic than the
best-fit SPEX model that measured a lower column, $N_{H} =
0.9^{+0.4}_{-0.4}\times 10^{24}~{\rm cm}^{-2}$.  Moreover, the XSPEC
model is simpler; it is constructed with fewer components.  On these
grounds, we suggest that the spectrum from ObsID 22885 was likely
obtained in a Compton-thick phase.  The XSPEC model for ObsID 22885
prefers a non-zero redshift for the reflector, nominally suggesting
that the reflecting gas is either infalling, located fairly close to
the central engine, or both.  This is consistent with
the implication of infall at the highest column densities in ObsID
22213; however, the best-fit model in SPEX does not require a
redshift.  The reflection scale factors differ considerably between
the XSPEC and SPEX models, owing to the different continua
assumed, and also the fact that the component is purely reflection
within XSPEC whereas the component carries a continuum within SPEX.

In Compton-thick spectra, it can be particularly difficult to
constrain the continuum and the column density.  The lower column
density implied in fits to the spectrum from ObsID 22886 is likely the
cause for a closer agreement in measurements of the internal
obscuration.  The best-fit XSPEC model measures a value of $N_{H} =
0.35^{+0.02}_{-0.02}\times 10^{24}~{\rm cm}^{-2}$; the best-fit SPEX
model measures $N_{H} = 0.29^{+0.04}_{-0.04}\times 10^{24}~{\rm
  cm}^{-2}$.  It is notable that the ``pion'' emission scale factor is
lower in this spectrum, $\Omega = 0.9^{+1.0}_{-0.5}$.  This suggests
that a lower emission scale factor corresponds to a lower obscuring
column, potentially indicating that the ionized re-emission and the
(nearly) neutral obscuration are connected despite having very
different gas properties.

Although the best-fit spectral models for ObsID 22285 and 22286 agree
on the observed flux in the 3--10~keV fitting band, they disagree
significantly on the implied unabsorbed flux and luminosity over the
0.5--30~keV band.  This is driven by the fact that the continuum is
not the same in the best-fit XSPEC and SPEX models; the ``comt''
component within SPEX goes to zero at both low energy and high energy,
whereas the cut-off power-law in XPSEC goes to infinity at low energy.
The fact that the ``refl'' component in SPEX includes a continuum
causes a degeneracy between its flux normalization and that of
``comt,'' leading to large fractional errors on these parameters.  

For both ObsID 22285 and 22286, we tested different formulations of
our basic model, aimed at exploring geometric departures from standard
AGN models.  Fits wherein the internal obscuration also acts on the
reflection component and/or the photoionized emission component are
rejected at more than the 5$\sigma$ level of confidence.  This
indicates that the emission lines are observed from far side of the
central engine, along a line of sight that intercepts little of the
diffuse gas on the near side.  Especially given that GRS 1915$+$105 is
viewed at a high inclination, this is only possible if the obscuration
is equatorial, consistent with disk winds (e.g., Miller et al.\ 2006,
Ponti et al.\ 2012).  ``Blurring'' the reflected emission and/or the
photoionized emission by the degree expected for Keplerian orbits at
$r = 10^{3-4}~GM/c^{2}$ is rejected by the spectra at the same level
of confidence.  This indicates that the reflection is observed farther
from the central engine than in ObsID 22213, and that the re-emission
is also distant.

Finally, we tested whether or not the neutral or low-ionization Fe~K
emission line in ObsIDs 22285 and 22286 is reflection, or if it may
arise in more diffuse gas.  A photon can lose a maximum of 150~eV per
Compton scattering event in cold, dense gas.  This leads to a
``Compton shoulder'' at 6.25~keV.  It may be made less distinctive by
Keplerian broadening, for instance, but this also acts on the narrow
line core and can be measured.  As noted above, significant broadening
is rejected by the data in ObsIDs 22285 and 22286.  Particularly in
ObsID 22286, which has slightly higher sensitivity than ObsID 2285,
the shoulder predicted by reflection is not clearly evident in the
data (see Figures 11 and 12).  This may be consistent with the most
sensitive spectra obtained from obscured AGN.  In {\it Chandra} HEG
spectra of NGC 1068, for instance, the Compton shoulder is also
absent, and Kallman et al.\ (2014) instead fit the line using a
low-ionization photoionized plasma component (the same ``photemis''
that we employed previously).

We therefore explored similar models for ObsIDs 22885 and 22886.  To
maintain contact with the results obtained from NGC 1068, we
constructed these models in XSPEC using ``photemis.''  If the
strongest line is not associated with neutral gas, then it is also
possible that the internal obscuration is not dominated by neutral
gas.  We constructed a model for each spectrum consisting of
(1) neutral line-of-sight obscuration, (2) ionized, potentially
partial-covering obscuration, (3) photoionized emission with a column
density and ionization parameter linked to the internal absorber, and
(4) photoionized emission from highly ionized gas.  The model
could be written as follows: $phabs\times (zxipcf*cutoffpl +
photemis_{1} + photemis_{2})$.  Like ``photemis,'' the ``zxipcf''
model is also built from executions of XSTAR, again enabling a level
of self-consistency.  The column density and ionization
of $zxipcf$ and $photemis_{1}$ were linked.  The parameters of
$zxipcf$ include the column density, ionization, covering fraction,
and redshift of the aborbing gas; we allowed all but the redshift to
float ($z=0$ was fixed in all fits).  Exploratory fits found that
solar abundances tended to over-predict the Si lines, so we fixed the
Fe abundance at twice the solar value to alter the Fe/Si ratio.  Prior
studies have suggested that the abundance of Fe may be elevated in GRS
1915$+$105 (e.g., Lee et al.\ 2002).

Fits to the spectrum obtained in ObsID 22285 achieve a Cash statistic
of $C = 198.9$ for 141 degrees of freedom (the fit is shown in Figure
13).  The low-ionization absorber and emitter was measured to have a
column density of $N_{H} \simeq 5.0\times 10^{23}~{\rm cm}^{-2}$, an
ionization of log$\xi = 1.1$, and a covering factor of $f=0.95$.  The
more ionized emission component had an ionization parameter of log$\xi
= 3.4$.  Both ``photemis'' components had a turbulent velocity of
$\sigma = 50~{\rm km}~{\rm s}^{-1}$.  The cut-off power-law component
had an index of $\Gamma = 1.8$, consistent with the lower-bound
enforced in our fits.

Fits to the spectrum from ObsID 22886 achieve a Cash statistic of $C =
650.8$ for 406 degrees of freedom.  In this case, the low-ionization
absorber and emitter was measured to have a column density of $N_{H} =
4.6\times 10^{23}~{\rm cm}^{-2}$, an ioniation of log$\xi = 1.59$, and
a covering factor of $f=0.95$.  The more highly ionized emitter had an
ionization parameter of log$\xi = 3.5$.  As with the fit to the
spectrum from ObsID 22885, the power-law index drifted to the lower
limit of the allowed range, $\Gamma = 1.8$.  Figure 14 shows that the
model still predicts Si emission lines that are stronger than
the data.  There is some residual flux in the Fe K band that is not
modeled; an additional photoionized emission zone with properties
intermediate between the low-ionization absorber and the highly
ionized absorber can account for some of this flux.

Clearly, as judged by the goodness-of-fit statistic, the fits achieved
using this alternative model were not as good as those achieved with
the more standard model.  However, there are potentially
considerations as important as the fit statistic.  A Compton shoulder
is a clear and unavoidable prediction of reflection from cold, dense
material.  The absence of such a shoulder in the spectra from GRS
1915$+$105 (and, e.g., NGC 1068) is potentially a matter of modest
sensitivity, but it could be meaningful.  If deeper {\it Chandra}
spectra and/or {\it XRISM} spectra reveal that the shoulder is truly
absent, then stratified wind models like ours may have to be adopted over
standard reflection models.

\section{Discussion}
We analyzed three high-resolution {\em Chandra}/HETG spectra of the
stellar-mass black hole GRS 1915$+$105.  The first observation was
obtained as the source entered a state with heavy internal
obscuration, and the latter two were made deep within this state.  One
of the latter observations likely occurred while the source was
enveloped by Compton-thick gas.  There is strong evidence that the
obscured state is the result of failed disk winds, originating
relatively close to the black hole and at a moderate Eddington
fraction.  This indicates that strong obscuration is not merely an
outcome of an [un]fortunate viewing angle, and not only seen in
super-Eddington sources like V404 Cyg (see, e.g., King et al.\ 2015,
Motta et al.\ 2017; also see Koljonen \& Tomsick 2020).  In this
section, we examine physical explanations for the failed wind in GRS
1915$+$105, address the consequences of our results for massive black
holes in Seyfert-2 and Compton-thick AGN, and note some unresolved
questions that can be addressed in future observations with {\it
  Chandra} and calorimeter spectrometers.

The low flux that is observed from GRS 1915$+$105 in the obscured
state is partly a matter of a reduced luminosity from the central
engine, and partly the result of heavy obscuration.  Especially when
obscuration becomes Compton-thick, it can be particularly difficult to
recover and constrain the true continuum spectrum in soft X-rays,
since only X-rays with $E\geq 20$~keV are able to penetrate the
obscuring gas (see, e.g., Kammoun et al.\ 2019, 2020).  Nevertheless,
deep within the obscured state, our modeling finds that the
0.5--30~keV luminosity of GRS 1915$+$105 ranged between $\lambda =
0.001-0.02$, where $\lambda = L/L_{Edd.}$.  This range is fully
consistent with that observed in Seyferts (see, e.g., Vasudevan \& Fabian 2009).

An examination of the plausible sub-Eddington wind driving mechanisms
clarifies why the flow in ObsID 22213 is a failed wind, and why any
flows in the latter observations are also unlikely to escape:

Radiative pressure on lines is only effective for log$\xi \leq 3$
(Stevens \& Kallman 1990, Proga 2003, Dannen et al.\ 2019).  Only the
inner zone is clearly blue-shifted, and its ionization is generally
above this limit.  Figure 8 shows a weak positive correlation between
ionization and outflow velocity, opposite to expectations if radiative
driving is operative.  Our analysis suggests that the wind launching
radius and ``reflection'' radius are comparable in ObsID 22213 ($few
\times 10^{3}~GM/c^{2}$); this may indicate that the ``reflection'' is
really re-emission from a very low-ionization component of the
wind. If radiation pressure acts on this component, the greater
quantity of ionized gas that would have to be dragged outward by
Coulomb forces likely prevents this mechanism from succeeding.

The wind originates much too close to the black hole for thermal
driving to succeed in launching it to infinity.  Using the disk
temperatures observed in unobscured stellar-mass black holes at
similar Eddington fractions as a guide (see, e.g., Reynolds et
al.\ 2013), $kT \simeq 0.3-1.0$~keV can be taken as the peak of the
overall spectrum and used as a proxy for the Compton temperature.
Then, a simple equation gives the radius at which thermal winds can be
launched: $R_{C} = 1.0\times 10^{10}~ (M_{BH}/M_{\odot}) / T_{C,8}$
(Begelman et al.\ 1983), or $R_{C} \simeq 1-3\times 10^{12}~{\rm cm}
\simeq 0.5-1.5\times 10^{6}~GM/c^{2}$ for GRS 1915$+$105.  Even if
thermal winds can be launched from $R\simeq 0.1~R_{C}$ as suggested by
Woods et al.\ (1996), this is still two orders of magnitude larger
than the upper limit on the inner absorption zone in ObsID 22213.  It
is possible, however, that the outer component of the wind is a
thermal flow.

Magnetic process can launch a disk wind from small radii.  Shakura \&
Sunyaev (1973) derived the magnetic field associated with the
$\alpha$-disk prescription as a function of $\alpha$, $\dot{m}$, and
$r$.  In the high/soft state of GRS 1915$+$105, Miller et al.\ (2016)
found that the expected field is at least an order of magnitude larger
than that required to launch and control the strong, fast winds
observed in that state.  Magnetic disk winds may therefore be natural
in the most luminous parts of the high/soft state, but this may not be
the case at lower $\dot{m}$.  If toroidal fields are to control the
gas flow, then the magnetic pressure must at least equal the gas
pressure: $B^{2}/8\pi \geq 2nkT$, or $B \geq \sqrt{16\pi nkT}$.  In
section 3.1, we estimated the density of the gas in the inner
absorption zone in ObsID 22213 to be $n\simeq 3.8\times 10^{14}~{\rm
  cm}^{-3}$.  This translates to $B \geq 3.3-5.8\times 10^{3}$~G,
depending on the assumed disk temperature.  Using the radii and
accretion rates that we have estimated, equation 2.19 in Shakura \&
Sunyaev (1973) predicts $B \leq 1.1\times 10^{4}$~G.  Thus, the
expected field is only comparable to that required in the obscured
state.  For plausible variations in the mass accretion rate and radius
of interest, then, the magnetic field in the disk may be unable to
expel the wind to infinity, resulting in a ``failed'' disk wind.

Figures 7 and 8 suggest positive correlations between the wind
parameters and luminosity, so it is possible that luminosity
variations in the disk affect the outermost part of the overall wind.
This could be part of a connected feedback cycle, since the mass flow
rate in the wind is roughly equivalent to that in the disk.  Shields
et al.\ (1986) examined instabilities and oscillations in thermal
winds; these are found to be important when the wind mass loss rate
exceeds the inflow rate by a factor of $\sim 15$.  The oscillation
periods predicted by Shields et al.\ (1986) are comparable to those
observed in GRS 1915$+$105, but the source appears to be an order of
magnitude below the critical $\dot{M}_{wind}/\dot{M}_{acc}$ threshold.
A recent numerical study by Ganguly \& Proga (2020) has found that
oscillations manifest at lower ratios, perhaps even at
$\dot{M}_{wind}/\dot{M}_{acc} \simeq 6$.  Our estimate of the loss
rate would approach this threshold if the wind was launched
vertically.  It is plausible that the observed oscillations are a
natural consequence of specific conditions and feedback in a failed
thermal wind.

It is also possible that the observed oscillations are fundamentally
magnetic.  Figure 8 shows that the flow is red-shifted at the highest
column densities, potentially when the gas density is highest.  These
phases would require the the strongest magnetic field to control and
launch the gas, and -- using the arguments above -- may exceed the
field that the disk can produce.  Thus, cycling across this pressure
balance could cause the variations observed in ObsID 22213.  

The general implication is that prolonged episodes of even standard
thin-disk accretion at a relatively low values of $\dot{m}$ may lead
to obscured states.  In X-ray binaries with shorter orbital periods,
smaller separations, and smaller disks, the time spent in the critical
combination of $\dot{m}$ and wind properties may be relatively short.
These sources may avoid lengthy obscured states.  However, the long
orbital period of GRS 1915$+$105 (${\rm P} = 33.85\pm 0.16$~days;
Steeghs et al.\ 2013) means that the same evolution may be relatively
slow, potentially leading to longer obscured states, or even leading
to the source becoming trapped in this state for several viscous time
scales.  Balakrishnan et al.\ (2020, in prep.) note that the viscous
time scale through the entire disk in GRS 1915$+$105 is likely about
500 days, and that the ``obscured state'' has (so far) only lasted
about that long.

One reason to examine stellar-mass black holes and AGN is that
although their inner accretion flows are likely to be similar, their
broader accretion flow structure may differ.  Especially in low-mass
X-ray binaries like GRS 1915$+$105, the outer disk is fed by Roche
lobe overlow -- there is no ``torus''.  In contrast, massive black
holes must be fed from gas that migrated into the nucleus, perhaps
settling into a torus-like geometry that eventually feeds a disk.  The
fact that the obscured state of GRS 1915$+$105 can be fit with models
that also describe Seyfert-2 and Compton-thick AGN signals that
obscuration in massive black holes need not arise in a very distant
torus.  This supports evidence of obscuration arising on scales
comparable to the optical broad line region in AGN, independently
inferred using variability (e.g., Elvis et al.\ 2004), and direct
spectroscopy (e.g., Costantini et al.\ 2007, Miller et al.\ 2018,
Zoghbi et al.\ 2019; also see Giustini \& Proga 2020).  Even strong
obscuration may be an accretion phenomenon defined in terms of
gravitational radii and $\dot{m}$ rather than physical units.

Our results also signal that episodes of high obscuration may
naturally manifest in AGN, at mass accretion rates typical of
Seyferts.  A fraction of highly obscured Seyferts may then simply be
in a normal evolutionary state.   This is likely consistent with
emerging evidence of a bimodality in the inclination of galaxies that
host Compton-thick AGN (Kammoun et al.\ 2020, in prep.).  If
obscuration was exclusively set by orientiation -- determined by a
torus-like structure that was fed by gas in the plane of the host
galaxy and therefore aligned with that plane -- then the distribution
of CTAGN should be smoothly distributed around $\theta = 60^{\circ}$
(the average viewing angle in three dimensions).

Currently, the HETGS is the only instrument that can measure the width
of the narrow lines in the obscured state of GRS 1915$+$105, and the
only instrument that could measure small velocity shifts.  A series of
monitoring observations with the HETGS would be able to determine
whether or not the neutral (or low-ionization) and ionized lines trace
the flux from the central engine, and to examine how the obscuration
varies with the luminosity of the central engine.  If monitoring
observations were to be spaced closely in time, it may even be
possible to search for variations in the obscuration and emission
lines as a function of the binary orbital phase.

The resolution of the {\it Resolve} calorimeter spectrometer aboard
{\it XRISM} is expected to be just 5~eV, approximately an order of
magnitude sharper than the HEG first order.  It will also have an
effective area of approximately $300~{\rm cm}^{2}$ in the Fe K band --
again, approximately an order of magnitude improvement over the HEG
first order (Tashiro et al.\ 2018).  With this resolution and
sensitivity, it will be possible to detect orbital broadening of the
re-emission spectrum from obscured states in stellar-mass black holes
and highly obscured AGN.  This will unambiguously determine the
location of the obscuring geometry.  The presence or absence of
Compton shoulders on the neutral or low-ionization Fe K line at
6.4~keV will also become clear, revealing whether the feature
originates in cold, dense gas or a more diffuse wind.  Finally, it
will be possible to determine the charge state of low-ionization Fe~K
emission lines.  In the longer run, the {\it X-IFU} calorimeter aboard
{\it Athena} is expected to achieve a resolution of 2.5~eV and to have
a collecting area roughly 100 times greater than the HEG first order
in the Fe K band (e.g., Barret et al.\ 2018); it will definitively
determine the nature and origin of obscuration in black holes across
the mass scale.

% Acknowledgments
We thank the {\it Chandra} Director, Belinda Wilkes, the mission
planning staff, and the HETG team for making these observations
possible.  JMM acknowledges helpful scientific conversations with
Keith Arnaud, Rob Fender, Richard Mushotzky, Jelle de Plaa, and Tahir
Yaqoob, computing assistance from Brandon Case, and the generosity of
Elizabeth Lauritsen Miller, Ivy Miller, and Ethan Miller (who made it
possible to finish this analysis during lockdown).

\clearpage

%----------------------------------------------------------------------

\begin{table}[t]
\caption{Observations}
\begin{footnotesize}
\begin{center}
\begin{tabular}{llll}
ObsID & Start time (UTD) & Start time (MJD) & Net Exposure (ks) \\
\tableline
\tableline
22213 & 2019-04-30 04:39:33 &   58603.19 & 29.1 \\
22885 & 2019-11-03 08:52:16 &   58790.37 & 29.3 \\
22886 & 2019-11-30 08:44:15 &   58817.36 & 29.4 \\
\tableline
\tableline
\end{tabular}
\vspace*{\baselineskip}~\\ \end{center} 
\tablecomments{{\it Chandra}/HETG exposures analyzed in this work.
  Please see the text for details of the instrumental set-up.}
\end{footnotesize}
\end{table}
\medskip

\begin{figure}
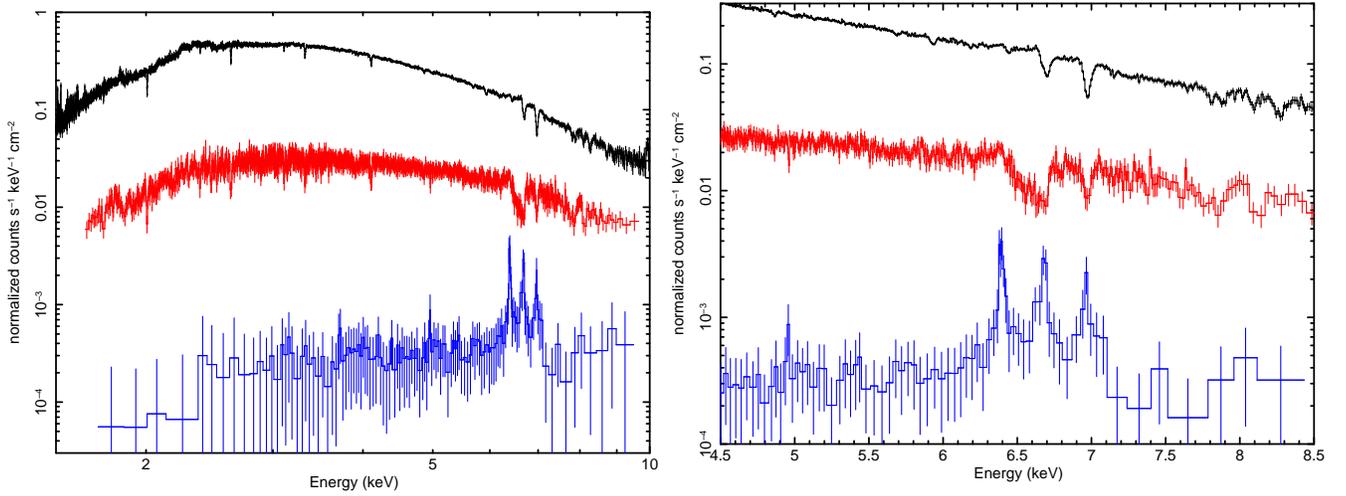

\hspace{-0.1in}
\includegraphics[scale=0.37,angle=-90]{f1a.ps}
\hspace{-0.25in}
\includegraphics[scale=0.37,angle=-90]{f1b.ps}
\figcaption[t]{\footnotesize GRS 1915$+$105 in states reminiscent of
  Seyfert 1 and Seyfert 2 AGN.  The lefthand panel shows the full pass
  band; the righthand panel focuses on the Fe K region.  In black: the
  spectrum from a steady soft state with a strong disk wind; accretion
  disk P-Cygni profiles are evident (ObsID 16711; Miller et
  al.\ 2016).  In red: the spectrum from ObsID 22213 as GRS 1915$+$105
  entered the obscured state.  The absorption trough between
  6.4--6.7~keV indicates a broad range of charge states.  In blue: the
  spectrum from ObsID 22885.  The emission lines are many times
  stronger than the local continuum.  The spectrum is likely
  consistent with the central source being embedded in Compton-thick
  obscuration.  (The spectra were been binned for visual clarity.)}
\vspace{0.25in}
\end{figure}
\medskip

\clearpage

\begin{figure}
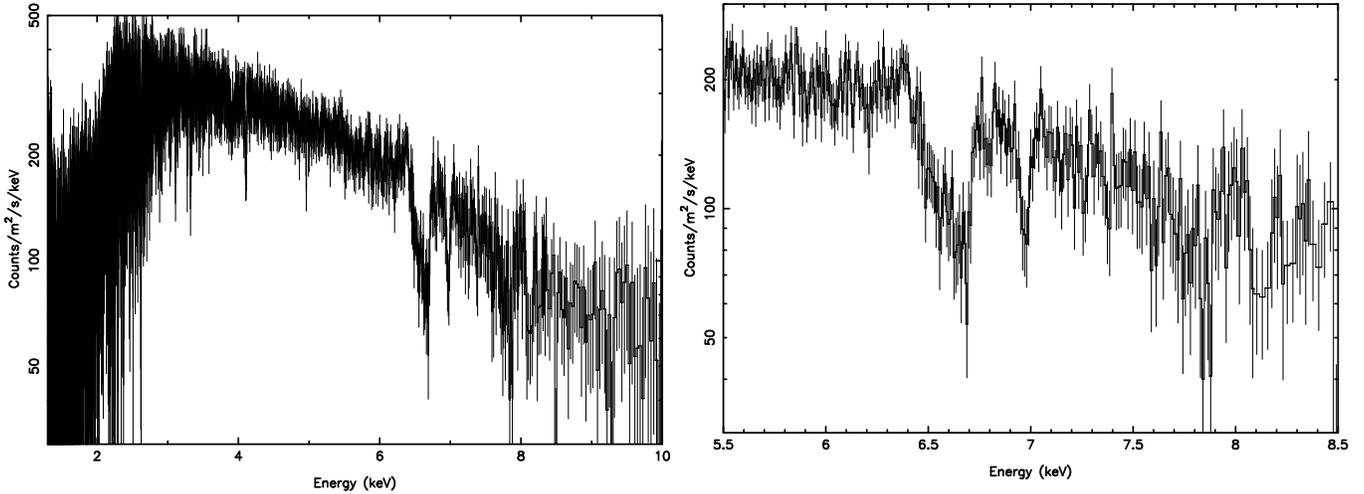

%\hspace{-0.1in}
\includegraphics[scale=0.36,angle=-90]{f2a.ps}
%\hspace{-0.25in}
\includegraphics[scale=0.36,angle=-90]{f2b.ps}
\figcaption[t]{\footnotesize The spectrum from ObsID 22213, obtained
  as GRS 1915$+$105 entered the obscured state (also see Figure 1).
  The data were binned within SPEX according to the ``optimal''
  binning scheme of Kaastra \& Bleeker (2016).  LEFT: The spectrum
  over the full pass band; He-like and H-like absorpotion lines form
  Si, S, Ar, Ca, and Fe are evident; lines from less abundant elements
  may also be present.  The Fe absorption features are particularly
  strong; the He-$\beta$ and Ly-$\beta$ lines of Fe are also evident.
  The only emission line is the neutral (or, low-ionization)
  Fe~K$\alpha$ line at 6.40~keV.  RIGHT: The same spectrum, in the Fe
  K band.  The broad absorption feature between 6.40--6.70~keV likely
  indicates that a range of Fe charge states likely contribute to the
  observed spectrum.  Initial fits confirmed that no single zone
  (e.g., no single ionization parameter) could account for this trough
  and the H-like line close to 6.97~keV.}
\vspace{0.25in}
\end{figure}
\medskip

\begin{figure}
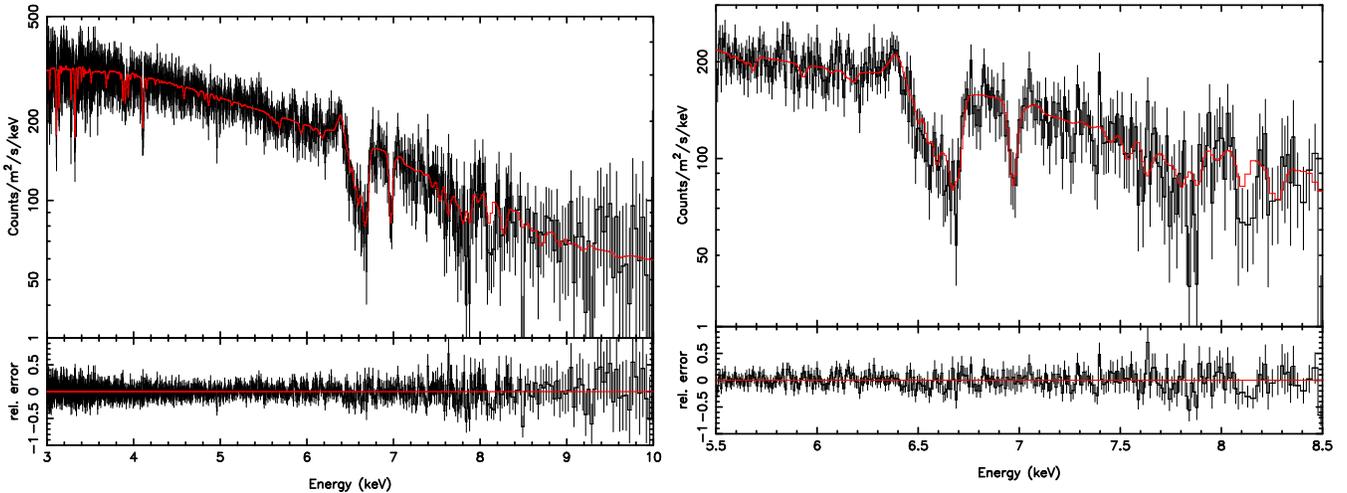

%\hspace{-0.1in}
\includegraphics[scale=0.36,angle=-90]{f3a.ps}
%\hspace{-0.25in}
\includegraphics[scale=0.36,angle=-90]{f3b.ps}
\figcaption[t]{\footnotesize The spectrum from ObsID 22213, captured
  as GRS 1915$+$105 entered the obscured state (also see Figure 1).
  The data were binned according to the ``optimal'' binning scheme of
  Kaastra \& Bleeker (2016).  The model
  (shown in red) included two layers of photoionized absorption to
  describe an apparent wind, and reflection with dynamical blurring.
  The wind is found to be dense, slow-moving, and to originate at
  small radii.  It may not escape from the system, and may build-up
  the obscuring material that later buries the central engine in a
  Compton-thick cloud.  LEFT: The spectrum and model on a broad pass
  band.  RIGHT: The spectrum and model in the Fe K band.  See Table 2
  for the spectral fit parameters.}
\vspace{0.25in}
\end{figure}
\medskip

\clearpage

\begin{figure}
%\hspace{-0.1in}
  \includegraphics[scale=0.9]{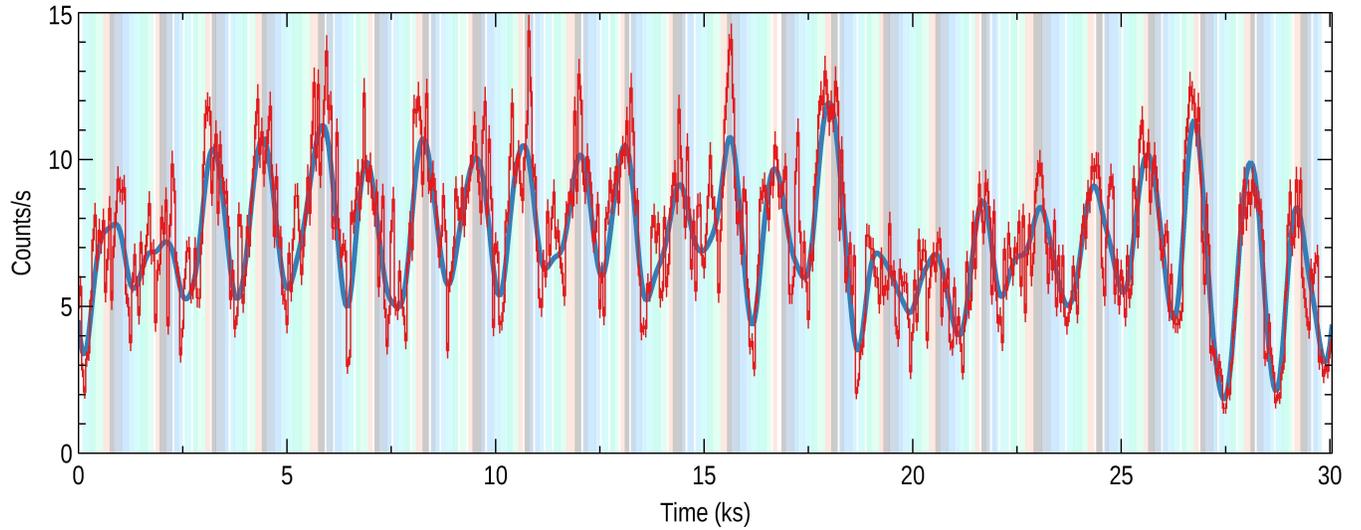}
%  \vspace{0.2in}
  \figcaption[t]{\footnotesize The light curve of the first-order HEG
    and MEG events in ObsID 22213, as GRS~1915$+$105 entered the
    obscured state.  The data are shown in red, and the light curve
    has been binned so that each point represents 50~s of integration.
    The variations have a quasi-period of approximately
    P$\simeq$1250~s, and their amplitude approaches $\pm$50\%.  The
    blue line is the result of Gaussian Process (GP; Rasmussen et
    al.\ 2005) modeling of the light curve (see the text for details).
    The GP results were used to define oscillation phase bins relative
    to the crests and troughs, and to then extract spectra from each
    bin (see Table 2).  The oscillation phase bins are indicated by
    the vertical stripes.}
\vspace{0.25in}
\end{figure}
\medskip

\clearpage

\begin{figure}[t!]
  \hspace{0.75in}
  \includegraphics[scale=0.80]{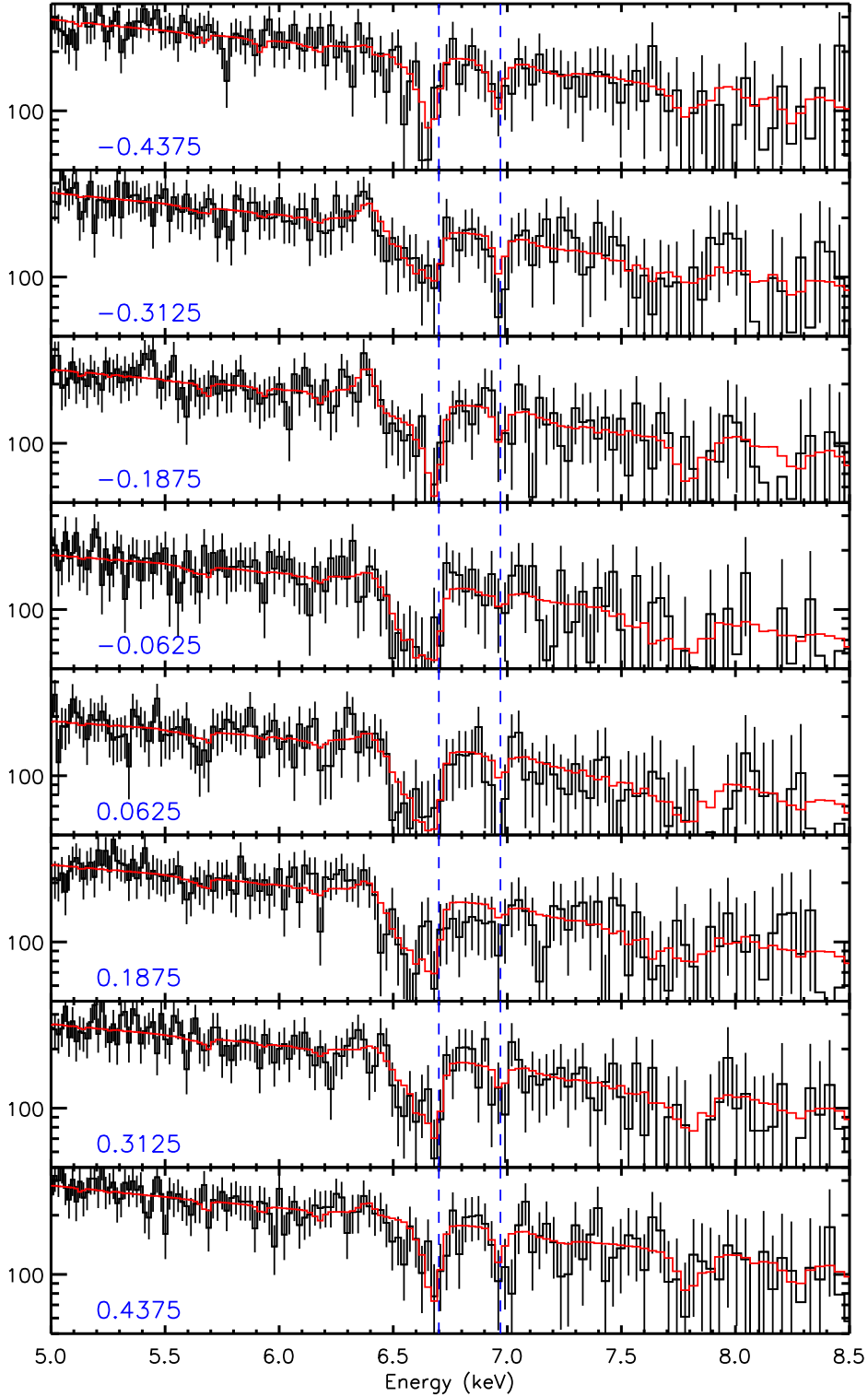}
\vspace{-0.5in}
  \figcaption[t]{\footnotesize Phase-selected spectra from ObsID
    22213, as GRS 1915$+$105 entered the obscured state.  The relative
    phase of each spectrum is indicated in blue.  The phase selections
    are shown in Figure 4, and the best fit for each phase is detailed
    in Table 2.  Dashed vertical blue lines indicate the rest energy
    of the He-like Fe XXV (6.70~keV) and H-like Fe XXVI (6.97~keV)
    resonance absorption lines.  The vertical axis in each panel is in
    units of ${\rm counts}~{\rm m}^{-2}~{\rm s}^{-1}~{\rm keV}^{-1}$.
    Each phase-selected spectrum has been fit with the same model used
    to fit the time-averaged spectrum.  The best-fit model for each
    phase-selected spectrum is shown in red.  Significant variations
    are seen in both of the photoionized absorbers and the reflector
    (again, see Table 2).}
\vspace{0.25in}
\end{figure}
\medskip

\clearpage

\begin{figure}[t!]
  \hspace{0.75in}
  \includegraphics[scale=0.75]{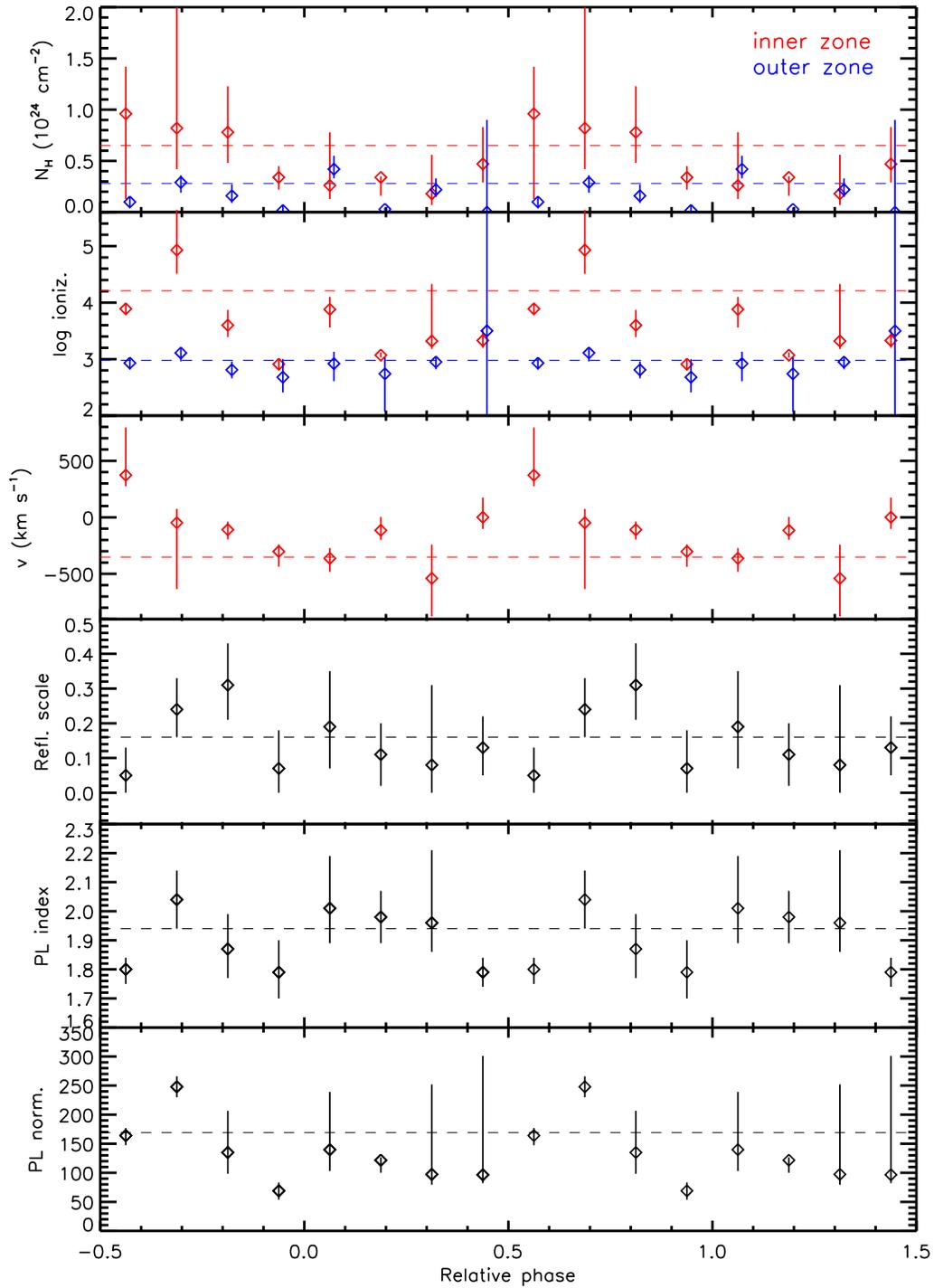}
\vspace{-0.5in}
  \figcaption[t]{\footnotesize Critical model parameters versus
    relative phase, based on spectral fits to phase-selected spectra
    from ObsID 22213 (see Figures 2--5, and Table 2).  Dashed
    horizontal lines in each panel indicate the best-fit value of the
    parameter measured in fits to the time-averaged spectrum.  The
    $1\sigma$ errors are plotted.  Two cycles are plotted for clarity.}
\vspace{0.25in}
\end{figure}
\medskip

\clearpage

\begin{figure}[t!]
  \hspace{0.5in}
  \includegraphics[scale=0.4]{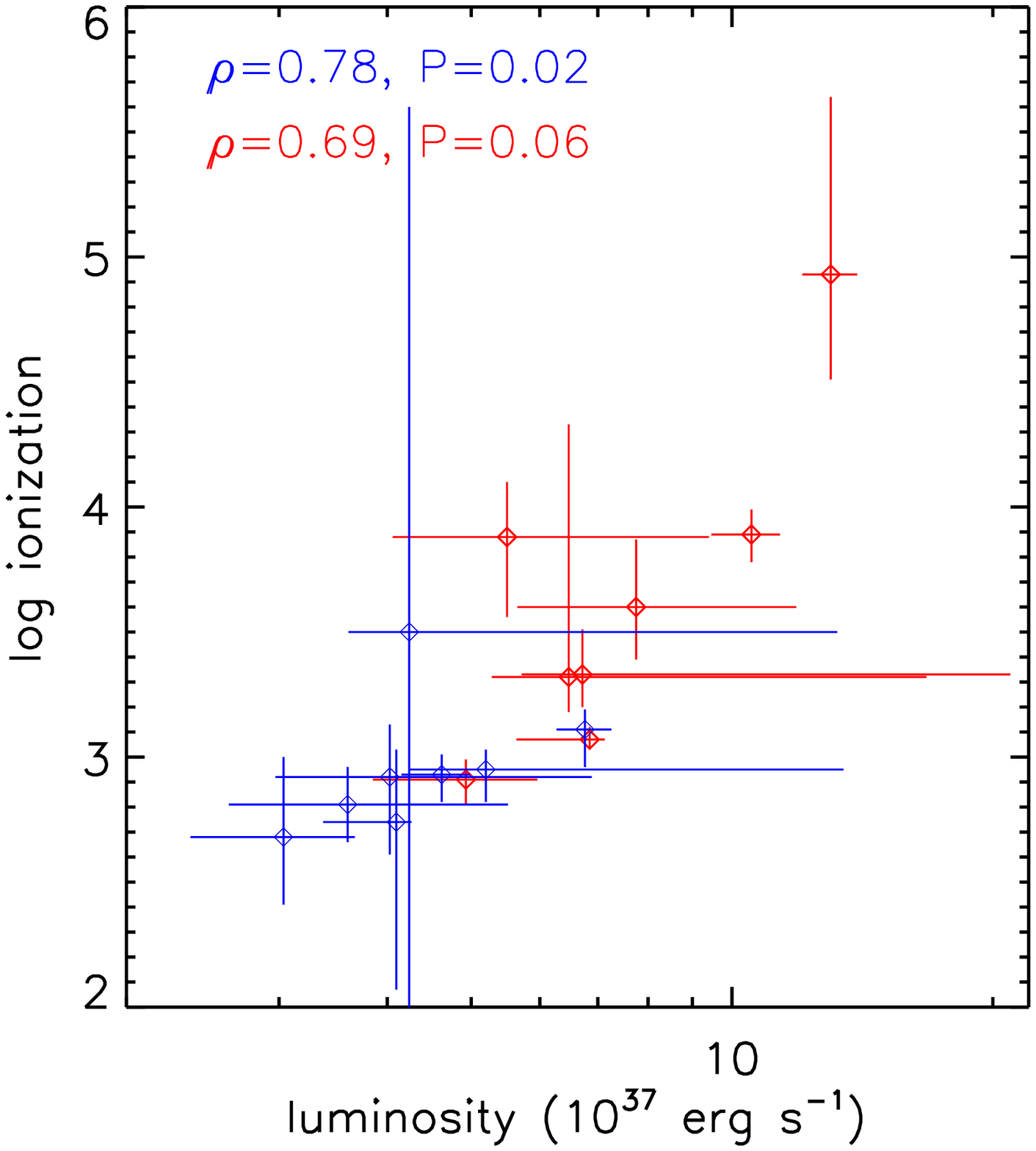}
  \hspace{-0.25in}
  \includegraphics[scale=0.4]{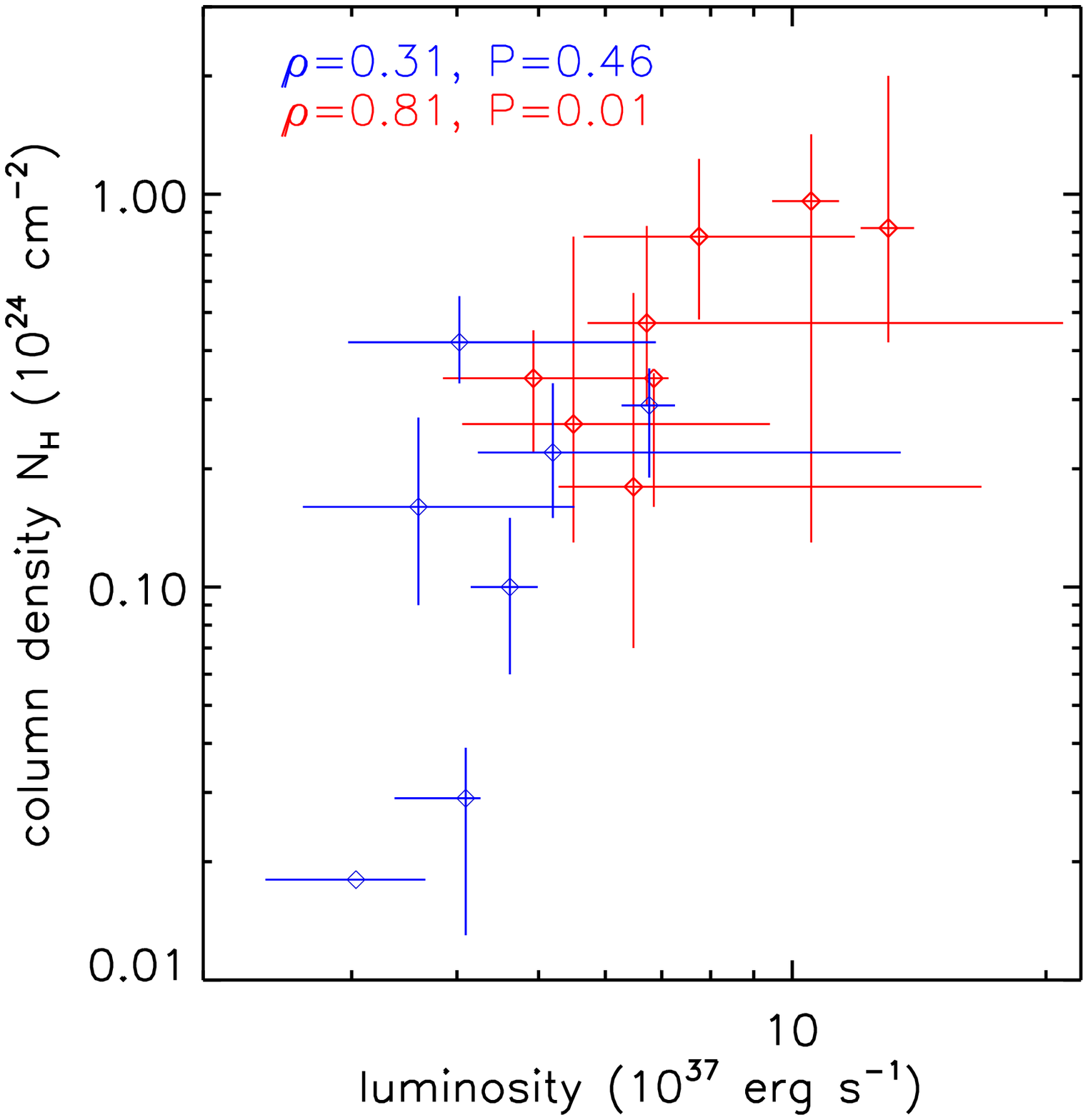}
%  \vspace{-0.5in}
  \figcaption[t]{\footnotesize Inner (red) and outer (blue)
    photoionized absorber properties versus luminosity, from fits to
    the ``phase-resolved'' spectra of ObsID 22213 as GRS~1915$+$105
    entered the obscured state (also see Table 2).  The Spearman's
    rank correlation coefficient, $\rho$, and the probability of false
    correlation, $P$, are noted for each zone in each panel.  The
    inner zone properties are correlated with luminosity; in contrast,
    the outer zone parameters are uncorrelated.  Note that the outer
    zone properties are evaluated and plotted versus the ``effective''
    luminosity seen by the zone, after the continuum has passed
    through the inner zone.}
\vspace{0.25in}
\end{figure}
\medskip

\begin{figure}[t!]
  \hspace{0.5in}
  \includegraphics[scale=0.4]{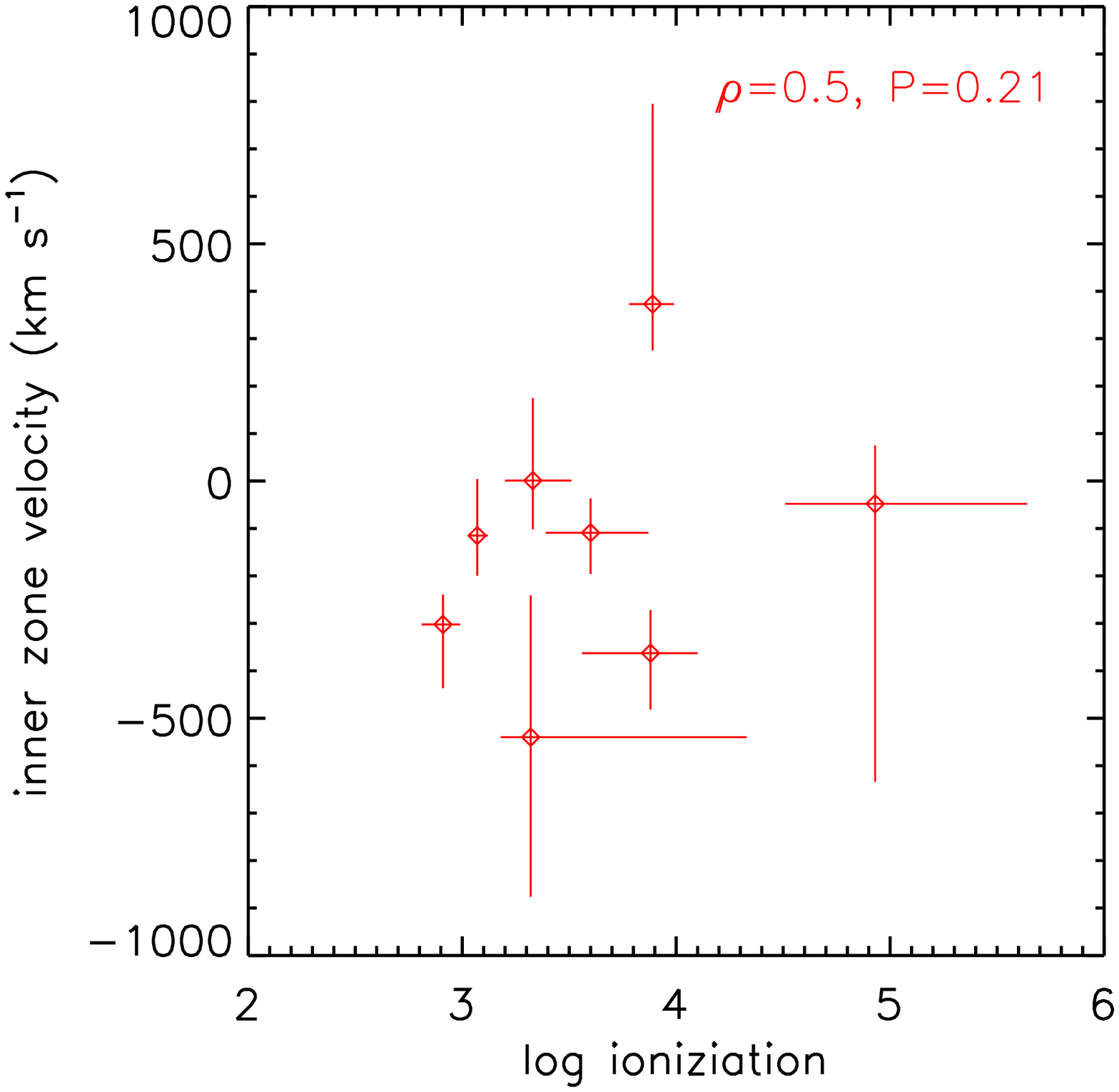}
  \hspace{-0.25in}
  \includegraphics[scale=0.4]{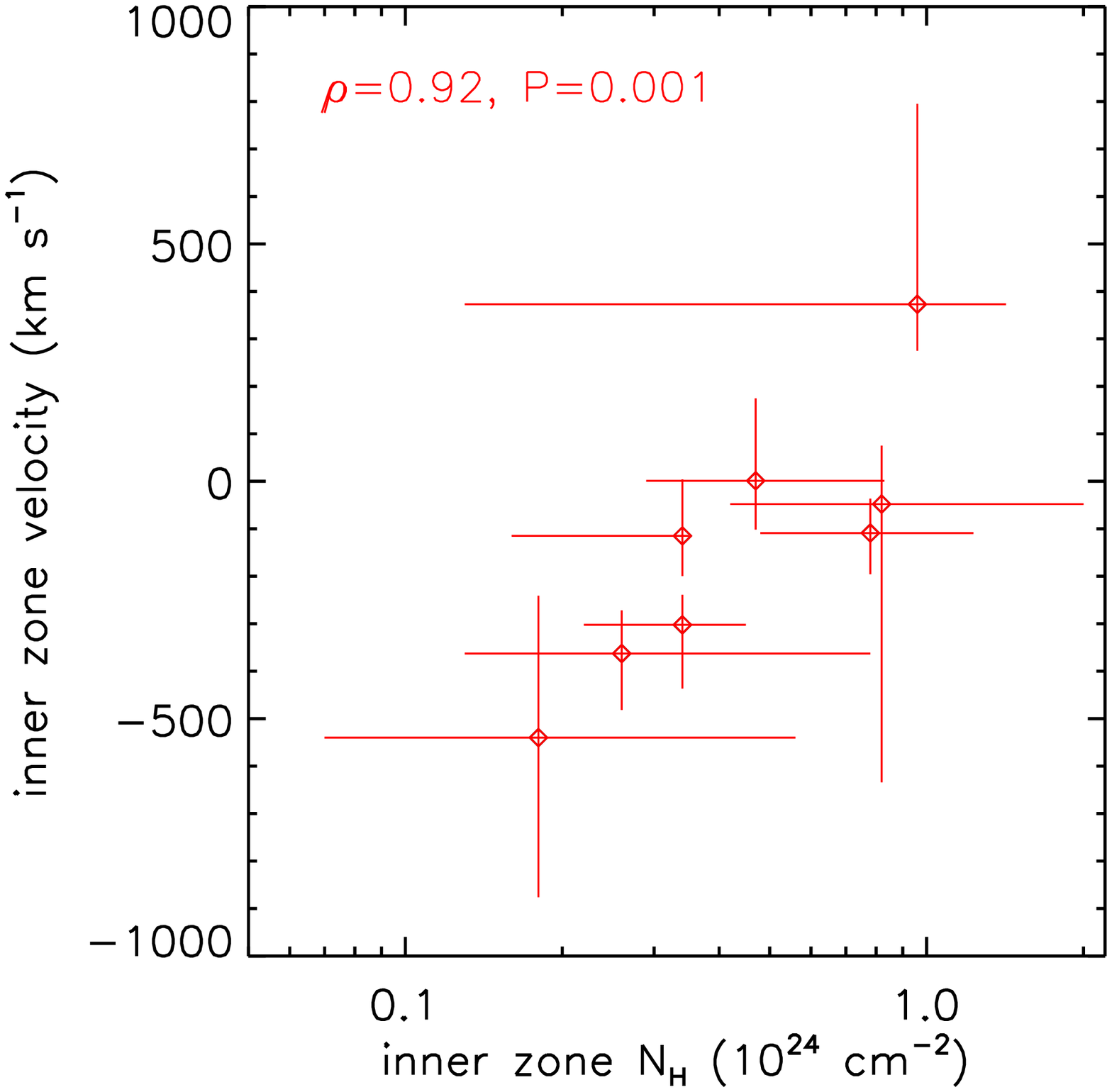}
%  \vspace{-0.5in}
  \figcaption[t]{\footnotesize Additional properties of the inner
    photoionization zone observed in ObsID 22213, as GRS 1915$+$105
    entered the obscured state.  Note that negative velocities are
    blue-shifts.  The Spearman's rank correlation coefficient, $\rho$,
    and the probability of false correlation, $P$, are noted in each
    panel.  The projected outflow velocity is not significantly correlated
    with the ionization of the gas, indicating that radiation driving
    is likely unimportant.  However, the outflow velocity is strongly
    and significantly correlated with the column density of the gas.
    {\em As the column density of the inner zone increases, the
      central engine is less able to expel the gas; at the highest
      observed columns, the gas is red-shifted.}  This may be an
    important clue as to the physical mechanisms that created the
    obscured state in GRS 1915$+$105.}
\vspace{0.25in}
\end{figure}
\medskip

\clearpage

\begin{table}[t]
\caption{Fits to ObsID 22213, as GRS 1915$+$105 Entered the Obscured State}
\begin{small}
\begin{center}
\begin{tabular}{llllllllll}
Parameter & Time-avg. & One & Two & Three & Four & Five & Six & Seven & Eight \\
     ~    &    ~      & $\phi=-0.4375$ & $\phi=-0.3125$ & $\phi=-0.1875$ & $\phi=-0.0625$ & $\phi=0.0625$ & $\phi=0.1875$  & $\phi=0.3125$ & $\phi=0.4375$ \\
\tableline
$N_{H,inner}$ & $0.7^{+0.1}_{-0.2}$ & $1.0^{+0.5}_{-0.8}$ & $0.8^{+2.7}_{-0.4}$ & $0.8^{+0.5}_{-0.3}$ & $0.3^{+0.1}_{-0.1}$ & $0.3^{+0.5}_{-0.1}$ & $0.34^{+0.01}_{-0.18}$ & $0.2^{+0.4}_{-0.1}$ & $0.5^{+0.4}_{-0.2}$ \\
log$\xi_{inner}$ & $4.2^{+0.1}_{-0.3}$ & $3.6^{+0.1}_{-0.1}$ & $4.9^{+0.7}_{-0.4}$ & $3.6^{+0.3}_{-0.2}$ & $2.9^{+0.1}_{-0.1}$ & $3.9^{+0.2}_{-0.3}$ & $3.07^{+0.05}_{-0.04}$ & $3.32^{+1.0}_{-0.1}$ & $3.3^{+0.2}_{-0.1}$ \\
$\sigma_{inner}$ & $70^{+18}_{-21}$	& $70^{*}$ & $70^{*}$ & $70^{*}$ & $70^{*}$ & $70^{*}$ & $70^{*}$ & $70^{*}$ & $70^{*}$ \\
$v_{inner}$ & $-35^{-70}_{+70}$ & $360^{+420}_{-100}$ & $-50^{-590}_{+120}$ & $-110^{-90}_{+70}$ & $-300^{-130}_{+60}$ & $-360^{-120}_{+90}$ &  $-120^{-90}_{+120}$ & $-540^{-3340}_{+300}$ & $0_{-100}^{+170}$ \\
\tableline
$N_{H,outer}$ & $0.28^{+0.05}_{-0.04}$ & $0.08^{+0.08}_{-0.04}$ & $0.29^{+0.07}_{-0.10}$ & $0.16^{+0.11}_{-0.07}$ & $0.018^{+0.05}_{-0.02}$ & $0.42^{+0.13}_{-0.09}$ &	$0.03^{+0.01}_{-0.02}$ & $0.2^{+0.1}_{-0.1}$ & $0.0^{+0.9}_{-0.0}$ \\
log$\xi_{outer}$ & $2.98^{+0.06}_{-0.06}$ &	$3.08^{+0.08}_{-0.11}$ & $3.11^{+0.08}_{-0.15}$ & $2.8^{+0.2}_{-0.2}$ & $2.7^{+0.3}_{-0.3}$ & $2.9^{+0.2}_{-0.1}$ & $2.7^{+0.3}_{-0.7}$ & $3.0^{+0.1}_{-0.1}$ & $5^{+5}_{-5}$ \\
$\sigma_{outer}$ & $30^{+10}_{-10}$ & $29^{*}$ & $29^{*}$ & $29^{*}$ & $29^{*}$ $29^{*}$ & $29^{*}$ & $29^{*}$ & $29^{*}$ \\
$v_{outer}$ & $-50^{-20}_{+40}$ & $0^{*}$ & $0^{*}$ & $0^{*}$ & $0^{*}$ & $0^{*}$ & $0^{*}$ & $0^{*}$ & $0^{*}$ \\
\tableline
$\Gamma$ & $1.94^{+0.05}_{-0.04}$ & $1.74^{+0.04}_{-0.05}$ & $2.0^{+0.1}_{-0.1}$ & $1.9^{+0.1}_{-0.1}$ & $1.8^{+0.1}_{-0.1}$ &	$2.0^{+0.2}_{-0.1}$ & 	$1.98^{+0.09}_{-0.07}$ & 	$2.0^{+0.3}_{-0.1}$ &	$1.79^{+0.05}_{-0.05}$ \\
scale & $0.16^{+0.04}_{-0.03}$ &	$0.05^{+0.08}_{-0.05}$ &	$0.24^{+0.09}_{-0.08}$ & 	$0.3^{+0.1}_{-0.1}$ &	$0.1^{+0.1}_{-0.1}$ & 	$0.2^{+0.2}_{-0.1}$ & 	$0.11^{+0.09}_{-0.09}$ &	$0.1^{+0.2}_{-0.1}$ & 	$0.13^{+0.0}_{-0.08}$ \\
$r_{inner}$ & $2200^{+7800}_{-1280}$ & $2200^{*}$ & $2200^{*}$ & $2200^{*}$ & $2200^{*}$ & $2200^{*}$ & $2200^{*}$ & $2200^{*}$ & $2200^{*}$ \\
cos($\theta$) & $0.95^{+0.01}_{-0.07}$ & $0.95^{*}$ & $0.95^{*}$ & $0.95^{*}$ & $0.95^{*}$ & $0.95^{*}$ & $0.95^{*}$ & $0.95^{*}$ & $0.95^{*}$ \\
Norm. & $170^{+40}_{-30}$ & $140^{+10}_{-20}$ & $250^{+20}_{-20}$ & $140^{+70}_{-40}$ & $70^{+10}_{-20}$ & $140^{+90}_{-40}$ &	$120^{+10}_{-20}$ & $100^{+1500}_{-20}$ & $100^{+290}_{-10}$ \\
\tableline
$F_{abs,3-10}$ & $1.2^{+0.3}_{-0.2}$ & $1.4^{+0.1}_{-0.1}$ & $1.26(9)$ & $1.1^{+0.6}_{-0.3}$ & $0.9(2)$ & $0.9^{+0.6}_{-0.3}$ & $1.14^{+0.05}_{-0.21}$ & $1.3^{+2.0}_{-0.2}$ &$ 1.4^{+4.2}_{-0.2}$ \\
% $F_{unabs,3-10}$ & $3.5^{+0.9}_{-0.6}$ & $4.0^{+0.3}_{-0.4}$ & $4.3(3)$ & $3^{+2}_{-1}$ & $1.8(4)$ & $2.0^{+1.4}_{-0.5}$ & $2.34^{+0.09}_{-0.42}$ & $2.3^{+3.7}_{-0.4}$ & $4.5^{+13.5}_{-0.7}$ \\
$F_{unabs,0.5-30}$ & $11.^{+3}_{-2}$ & $11.9^{+0.9}_{-1.3}$ & $15(1)$ & $9^{+4}_{-2}$ & $6(1)$ & $6.2^{+4.4}_{-1.6}$ & $7.7^{+0.3}_{-1.4}$ & $7.3^{+11.7}_{-1.3}$ & $7.6^{+22.8}_{-1.2}$ \\
$L_{0.5-30}$ & $10^{+3}_{-2}$ & $10.6^{+0.8}_{-1.2}$ & $13.1(9)$ & $8^{+4}_{-2}$ & $5(1)$ & $5.5^{+3.9}_{-1.4}$ & $6.9^{+0.3}_{-1.2}$ & $6.5^{+10.4}_{-1.2}$ & $6.7^{+20.3}_{-1.0}$ \\

\tableline
C-stat. & 1148.8 & 424.5 & 390.9 & 401.1 & 439.3 & 407.0 & 397.0 & 386.6 & 359.9 \\
$\nu$ (dof) & 1090 & 423 & 378 & 378 & 378 & 378 & 378 & 378  & 378 \\
\tableline
\end{tabular}
\vspace*{\baselineskip}~\\ \end{center} 
\tablecomments{Fit parameters to the time-averaged and phase-selected
  spectra of ObsID 22213, grouped by inner photoionization zone, outer
  photoionization zone, reflection parameters, and fit statistics.
  The fits were made in SPEX using an overall line of sight column
  density ($N_{H} = 5.3\times 10^{22}~{\rm cm}^{-2}$, fixed in all
  cases), two photionized absorption zones (via the ``pion'' model),
  and reflection (via ``refl'').  Please see the text for details.
  Column densities are in units of $10^{24}~{\rm cm}^{-2}$.  The
  $\sigma$ parameter is the rms broadening within each photoionization
  zone in units of ${\rm km}~{\rm s}^{-1}$; the time-averaged value
  was fixed in fits to the phase-selected spectra.  The $v$ parameter
  is the motion of the zone with respect to the line of sight in units
  of ${\rm km}~{\rm s}^{-1}$; negative velocities are blue-shifts.
  The velocity of the outer zone is consistent with zero in the
  time-averaged spectrum and poorly constrained in the phase-selected
  spectra, so a value of zero was fixed in fits to the phase-selected
  spectra.  The reflection ``scale'' parameter is similar to the
  ``reflection fraction'' in other models.  The $r_{inner}$ parameter
  is the inner ``blurring'' radius for the reflection component; it
  was possible to constrain this parameter in fits to the
  time-averaged spectrum, and this value was then fixed in fits to the
  phase-selected spectra.  The cos($\theta$) parameter is the angle at
  which the reflector is viewed with respect to the line of sight;
  this parameter could only be constrained in the time-averaged
  spectrum and was then fixed in fits to the phase-selected spectra.
  The reflection normalization is in units of $10^{44}~{\rm ph}~{\rm
    s}^{-1}~{\rm keV}$.  Flux values are quoted in the 3--10~keV band,
  as observed (absorbed) in units of $10^{-9}~{\rm erg}~{\rm
    cm}^{-2}~{\rm s}^{-1}$.  Unabsorbed flux values are also quoted
  for an extrapolation to the 0.5--30~keV band; luminosities in this
  band are quoted in units of $10^{37}~{\rm erg}~{\rm s}^{-1}$.  Note
  that the time-averaged luminosity corresponds to an Eddington
  fraction of $\lambda = 0.06$ although the observed flux is
  fractionally much lower than typical values.}
\end{small}
\end{table}
\medskip

\clearpage

\begin{table}[t]
  \caption{Fits to ObsIDs 22885 and 22886, Deep within the Obscured State}
\begin{footnotesize}
\begin{center}
\begin{tabular}{llllll}
  Parameter &    22885   &   22885    &    22886   &    22886   \\
  ~         & (xillver$+$photemis) &  (refl$+$pion)  &  (xillver$+$photemis)  &  (refl$+$pion) \\
\tableline
\tableline
$N_{H} (10^{24}~{\rm cm}^{-2})$ & $2.2^{+0.9}_{-0.3}$ & 0.9(4) & 0.35(2) & 0.29(4) \\

\tableline

$\Gamma$ & $2.20_{-0.03}$ & --  & $2.20_{-0.02}$ & -- \\
$\tau$ & -- & 0.65(7) & -- & 0.6(2) \\
$K_{continuum}$ & $0.7^{+2.0}_{-0.3}$ & $0.03^{+0.02}_{-0.01}$ & 0.12(1) & 0.09(2) \\
$z (10^{-3})$ & 2.5(5) & -- & 1.5(6) & -- \\

\tableline

$f$ or $scale$ & $-0.09^{-0.01}_{+0.02}$ & $3^{+2}_{-1}$ & $-0.14(3)$ & $1.7(2)$ \\
$K_{reflection}$ & $0.024(2)$ & $2(1)$ & $0.037(3)$ & $5.3(2)$ \\

\tableline

$N_{H,emis} (10^{24}~{\rm cm}^{-2})$ & -- & $0.12^{+0.15}_{-0.06}$ & -- & $0.8^{+1.5}_{-0.5}$ \\

log$\xi_{emis}$ & $3.40(3)$ & $2.6(2)$ & $3.50(2)$ & $2.7^{+0.2}_{-0.1}$ \\
$z (10^{-3})$ & $0.0^{+0.2}$ & -- & $0.0^{+0.2}$ & -- \\
$K_{emis} (10^{2})$ & $2.5(3)$ & -- & $5.8^{+0.6}_{-0.3}$ &  -- \\
$\Omega$ & -- & $5^{+4}_{-1}$ & -- & $0.9^{+1.0}_{-0.5}$ \\

\tableline

$F_{abs,3-10}$ &  $3^{+8}_{-1}$  &  $3^{+2}_{-1}$    &    $8.8(7)$  &   $9(2)$  \\

$F_{unabs,0.5-30}$ & $3^{+8}_{-1}$ & $5^{+3}_{-2}$ & $81(7)$ & $21(4)$    \\

$L_{0.5-30}$ &   $3^{+8}_{-1}$  &  $0.44^{+0.3}_{-0.1}$  & $0.72(6)$    &  $0.19(4)$ \\

% The models get the unabsorbed flux to be the same, the rest is speculation.

\tableline  

$C$-stat & $154.8$ & $224.5$ & $530.0$ & $233.8$ \\
$\nu$ (dof)  & 140 & 192 & 414 & 187 \\

\tableline
\tableline
\end{tabular}
\vspace*{\baselineskip}~\\ \end{center} 
\tablecomments{Best-fit parameters, errors, and goodness-of-fit
  statistics from independent models for ObsIDs 22885 and 22886,
  obtained deep in the obscured state.  The models also included
  line-of-sight absorption within the Milky Way, fixed at a value of
  $N_{H} = 5.0\times 10^{22}~{\rm cm}^{-2}$ in all cases.  The spectra
  were binned slightly differently when fitting in XSPEC and SPEX,
  leading to different numbers of bins and degrees of freedom.  Within
  XSPEC, each observation was fit via $phabs_{los} \times (phabs\times
  cutoffpl + xillver + photemis)$.  For the $xillver+photemis$ model
  within in XSPEC, $phabs_{los}$ and $phabs$ components described
  line-of-sight and internal obscuration, respectively, the $cutoffpl$
  describes the obscured direct continuum (characterized by a photon
  indix $\Gamma$ and normalization), $xillver$ describes the neutral
  reflected emission (continuum parameters were linked to those in
  $cutoffpl$ where possible, others were fixed, leaving the redshift
  $z$, reflection fraction $f$, and normalization as free parameters),
  and $photemis$ describes photoionized emission lines (free
  parameters included the ionization parameter $\xi$, the redshift of
  the emission, and the normalization of the emission).  For the
  $refl+pion$ model within SPEX, each specrum was fit via
  $absm_{los}\times (absm\times compt + refl + pion)$.  The $absm$
  component is the same as $phabs$ within XSPEC, $compt$ is a simple
  Comptonization model (several parameters were fixed, leaving the
  optical depth $\tau$ and normalization as free parameters), $refl$
  is a reflection model based on $pexrav$ (the free parameters
  included the reflection scale factor and normalization), and pion is
  a self-consistent photoionized emission model that adjusts the
  ionization balance as the direct continuum adjusts within the fit
  (free parameters included the column density $N_{H}$, the ionization
  paramter $\xi$, and the covering factor $\Omega$).  The observed
  flux in the 3--10~keV fitting band is given in units of
  $10^{-11}~{\rm erg}~{\rm cm}^{-2}~{\rm s}^{-1}$.  The ``unabsorbed''
  flux in the 0.5--30~keV band is given in units of $10^{-9}~{\rm erg}~{\rm cm}^{-2}~{\rm s}^{-1}$.  The luminosity values in that band are given in units of 
$10^{37}~{\rm erg}~{\rm s}^{-1}$.  Please see the
  text for additional important details.}
\end{footnotesize}
\end{table}
\medskip

\clearpage

\begin{figure}
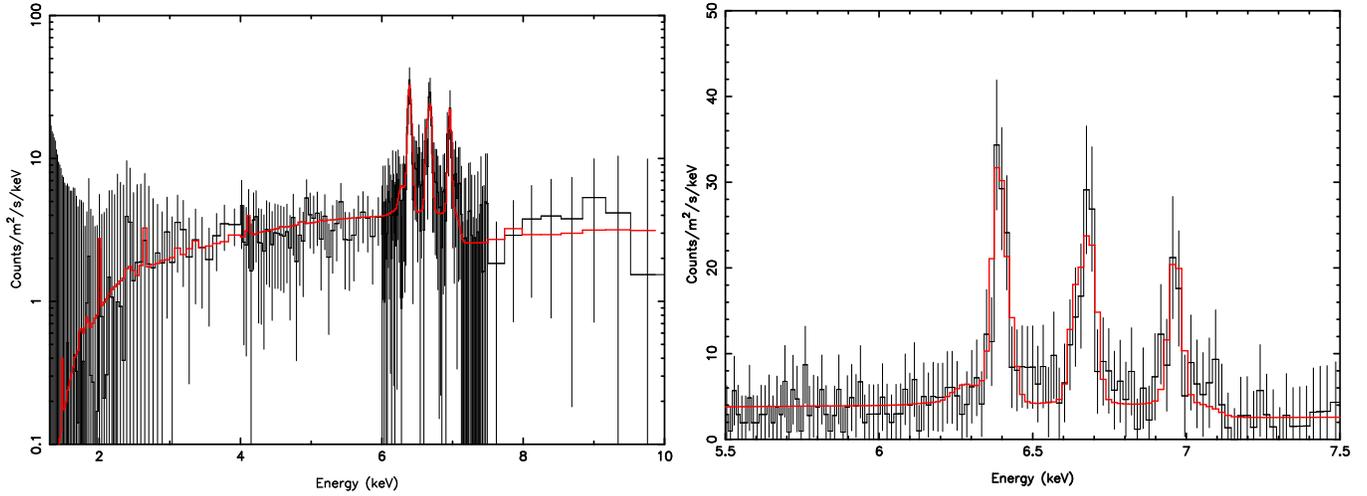

\includegraphics[scale=0.36,angle=-90]{f9a.ps}
\includegraphics[scale=0.36,angle=-90]{f9b.ps}
\figcaption[t]{\footnotesize The combined first-order {\it
    Chandra}/HETG spectrum of GRS 1915$+$105 in ObsID 22885.  The
  spectrum was fit over the broad pass band shown in the lefthand
  panel; the righthand panel focuses on the crucial Fe K band.  Note
  that the neutral Fe~K$\alpha$ line at 6.40~keV, the He-like complex
  at 6.70~keV, and H-like line at 6.97~keV are all several times
  stronger than the local continuum.  This indicates that most of the
  ionizing continuum is obscured from direct view.  The fits shown
  here were made in SPEX using a model consisting of line-of-sight
  absorption, absorption within GRS 1915$+$105 acting on a thermal
  Comptonizaton and linked, self-consistent photioizonized emission
  component (via ``pion''), and distant neutral reflection without
  internal obscuration (via ``refl'', based on ``pexmon'').  See Table
  3 for best-fit parameter values and errors.}
\vspace{0.25in}
\end{figure}
\medskip

\begin{figure}
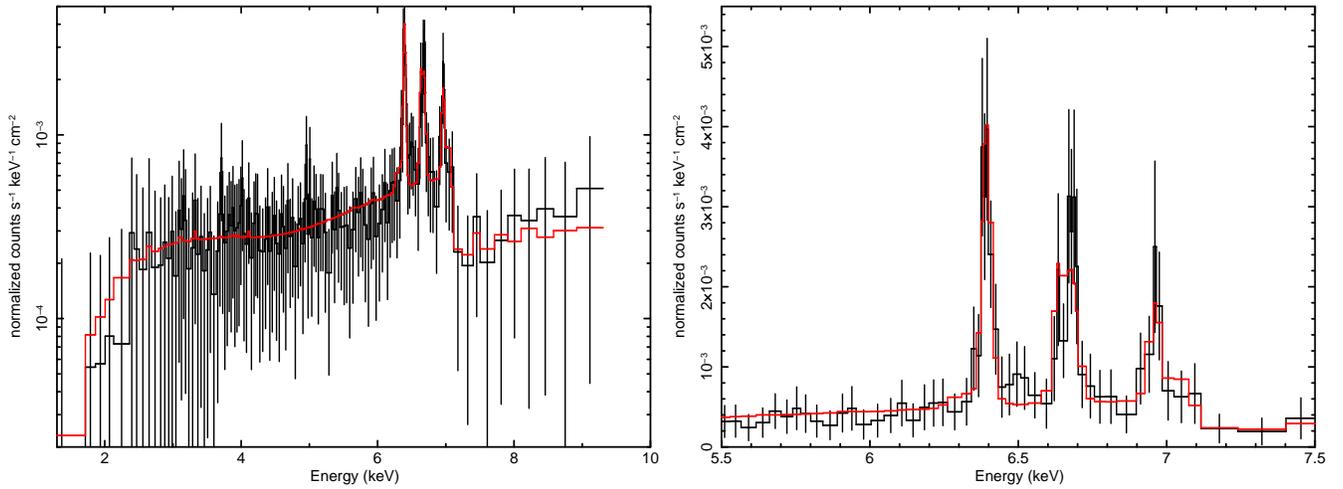

\hspace{-0.1in}  
\includegraphics[scale=0.37,angle=-90]{f10a.ps}
\hspace{-0.25in}
\includegraphics[scale=0.37,angle=-90]{f10b.ps}
\figcaption[t]{\footnotesize Similar to Figure 9, but in this figure
  the spectrum of ObsID 22885 was fit within XSPEC.  The model
  consists of line-of-sight absorption, absorption within GRS
  1915$+$105 acting on a cut-off power-law and a photioizonized
  emission component (via the XSTAR-derived ``photemis''), and distant
  neutral reflection without internal obscuration (via ``pexmon'').
  The spectrum consistent with being Compton-thick (see Table 3).  A
  slightly different binning is used in these fits; please see the
  text for details.
}
\vspace{0.25in}
\end{figure}
\medskip

\clearpage

\begin{figure}
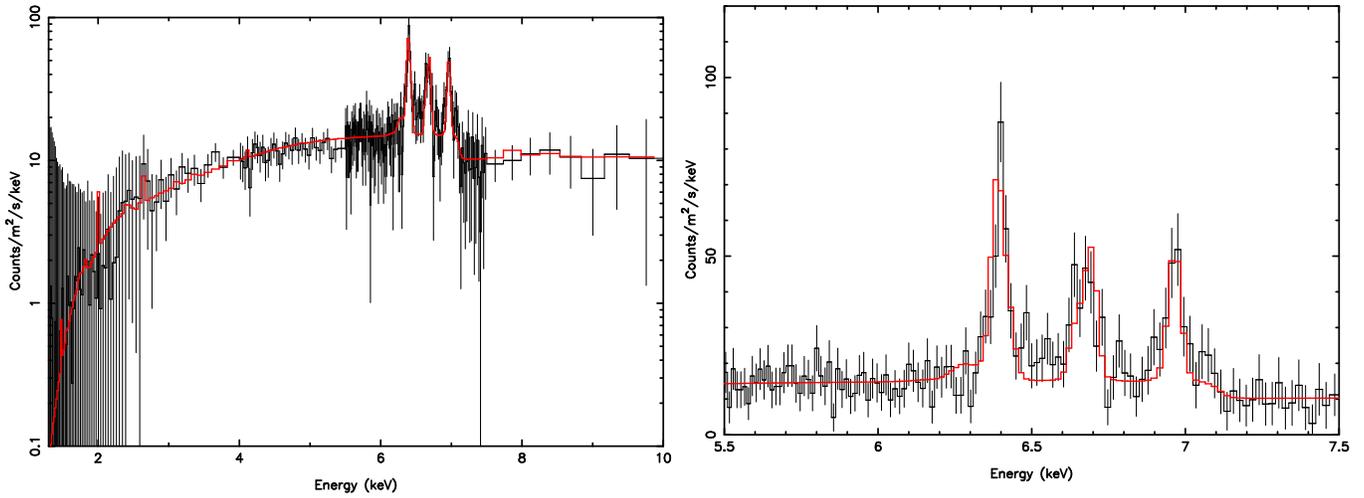

\includegraphics[scale=0.36,angle=-90]{f11a.ps}
\includegraphics[scale=0.36,angle=-90]{f11b.ps}
\figcaption[t]{\footnotesize Similar to Figure 9, but the spectrum
  from ObsID 22886 and corresponding SPEX model are shown.  The
  best-fit model suggests strong obscuration that is not quite
  Compton-thick (see Table 3).
}
\vspace{0.25in}
\end{figure}
\medskip

\begin{figure}
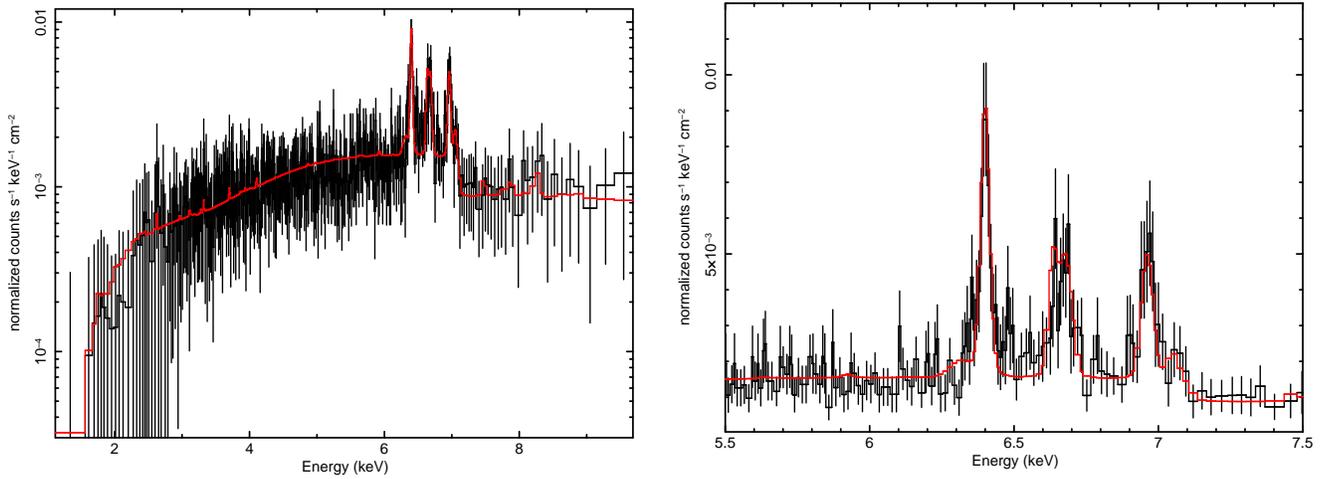

\includegraphics[scale=0.36,angle=-90]{f12a.ps}
\includegraphics[scale=0.36,angle=-90]{f12b.ps}
\figcaption[t]{\footnotesize Similar to Figure 11, but the XSPEC model
  for ObsID 22886 is shown.  This model again suggests strong
  obscuration that is not quite Compton-thick (see Table 3).  This is
  crudely indicated by the somewhat shallower depth of the Fe K edge
  in this observation, relative to ObsID 22885 (see Figures 9--10).
  In this spectrum and with the binning adopted here, the
  contributions of the He-like triplet components are relatively
  clear; the strong forbidden line is a clear indication of
  photoionzation.}
\vspace{0.25in}
\end{figure}
\medskip

%----------------------------------------------------------------------

\begin{figure}
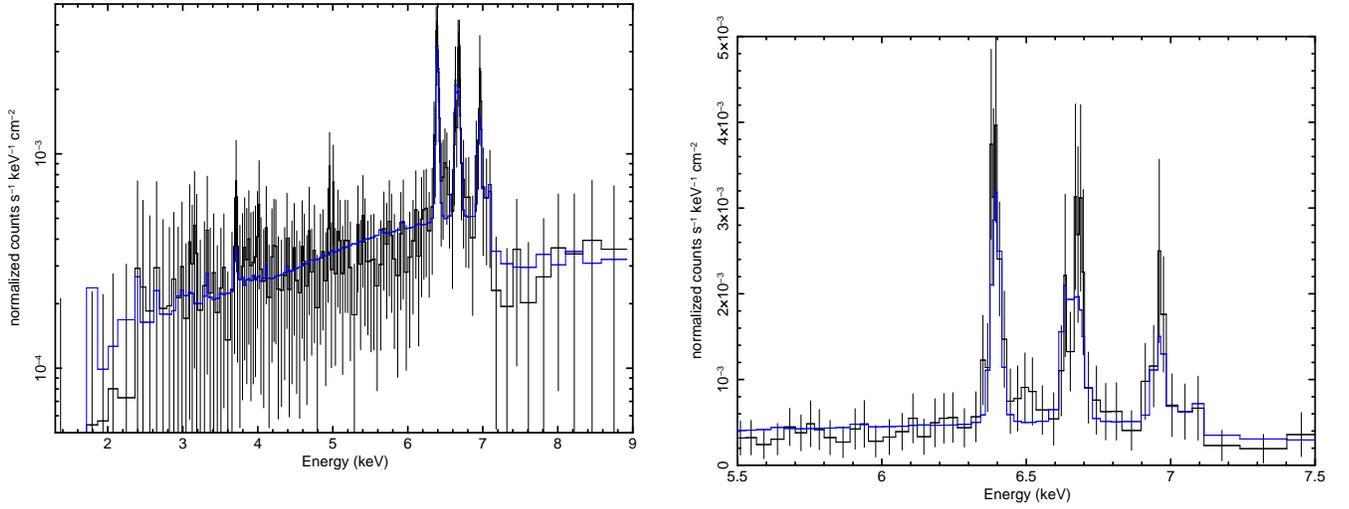

\includegraphics[scale=0.36,angle=-90]{alt22885_broad.ps}
\includegraphics[scale=0.36,angle=-90]{alt22885_narrow.ps}
\figcaption[t]{\footnotesize Fits to ObsID 22885 with an alternative
  model.  The 6.4~keV line is no longer fit with neutral reflection,
  and the internal obscruation is no longer fit with neutral
  absorption.  Instead, photoionized gas with a low ionization
  parameter (log$\xi\simeq 1.1$) is used to self-consistently model
  both the heavy internal obscuration and the 6.40~keV emission line.
  The absorber has a column density of $N_{H} \simeq 5\times
  10^{23}~{\rm cm}^{-2}$ and a covering fraction of $f\simeq0.95$.
  The He-like and H-like Fe lines are fit with a separate photoionized
  emission component (log$\xi\simeq 3.4$).  The overall fit is not as
  good as the fit that relied on reflection ($C-stat = 199$ versus
  $C-stat = 155$), but the model does not predict a Compton shoulder
  on the Fe K line and further improvements to the model may make it
  competitive with the more standard picture.  Please see the text for additional details and comments.}
\vspace{0.25in}
\end{figure}
\medskip

\begin{figure}
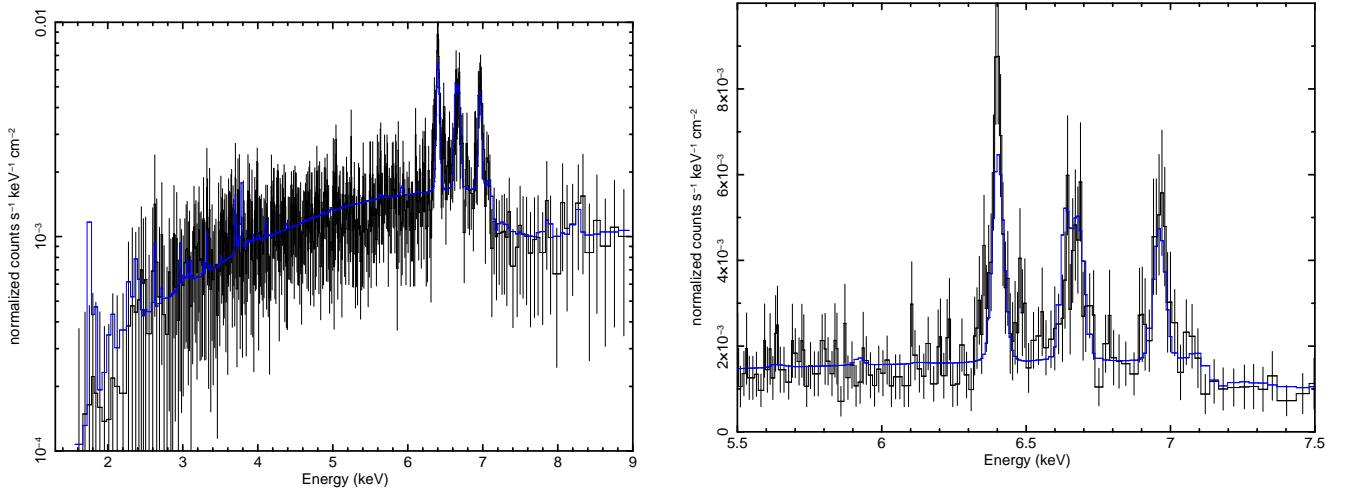

\includegraphics[scale=0.36,angle=-90]{alt22886_broad.ps}
\includegraphics[scale=0.36,angle=-90]{alt22886_narrow.ps}
\figcaption[t]{\footnotesize Similar to Figure 13, but here ObsID
  22886 is fit with the alternative model.  The coupled low-ionization
  absorber and re-emitter has a column density of $N_{H} \simeq
  5\times 10^{23}~{\rm cm}^{-2}$, a covering fraction of $f\simeq
  0.95$, and an ionizatin paramter of log$\xi\simeq 1.3$.  The more
  ionized photoionized zone is characterized by log$\xi\simeq 3.5$.
  The overall fit is not as good as the fit that relied on reflection
  ($C-stat = 580$ versus $C-stat=560$), but the model does not predict
  a Compton shoulder on the Fe K line and further improvements to the
  model may make it competitive with the more standard picture.
  Please see the text for additional details and comments.}
\vspace{0.25in}
\end{figure}
\medskip

\end{document}